\documentclass[12pt,preprint]{aastex}  

\begin{document}

\title{Population III star formation in a $\Lambda$CDM universe, \
I:  The effect of formation redshift and environment on protostellar accretion rate}
\author{Brian W. O'Shea\altaffilmark{1} \& Michael L. Norman\altaffilmark{2}}

\altaffiltext{1}{Theoretical Astrophysics (T-6), Los Alamos National
Laboratory, Los Alamos, NM 87545; bwoshea@lanl.gov}

\altaffiltext{2}{Center for Astrophysics and Space Sciences,
University of California at San Diego, La Jolla, CA 
92093; mnorman@cosmos.ucsd.edu}

\begin{abstract}
We perform 12 extremely high resolution adaptive mesh refinement
(AMR) cosmological hydrodynamic simulations of Population III star 
formation in a $\Lambda$CDM universe, varying the box size and
large-scale structure, to understand systematic effects in the formation of
primordial protostellar cores.  We find results that are qualitatively similar to
those of \markcite{ABN02}{Abel}, {Bryan}, \& {Norman} (2002), \markcite{2002ApJ...564...23B}{Bromm}, {Coppi}, \&  {Larson} (2002), and \markcite{2006astro.ph..6106Y}{Yoshida} {et~al.} (2006).
Our calculations indicate that the halos out of which these stars form have similar
spin parameters to galaxy-mass halos which form at $z \sim 0$
as well as with previous simulations of high-redshift halos.  
We observe that, in the absence of a photodissociating ultraviolet background, the 
threshold halo mass for formation of a Population III protostar does not evolve 
significantly with time in the redshift range studied ($33 > z > 19$) but exhibits
substantial scatter ($1.5 <$~M$_{vir}/10^5$~M$_\odot < 7$) due to different
halo assembly histories:  Halos which assembled more slowly develop
cooling cores at lower mass than those that assemble more rapidly, in 
agreement with~\markcite{2003ApJ...592..645Y}{Yoshida} {et~al.} (2003).  
We do, however, observe significant evolution in the 
accretion rates of Population III protostars with redshift, with objects that form 
later having higher maximum accretion rates ($\dot{m} \simeq 10^{-4}$~M$_\odot$/yr 
at $z=33$ and $\simeq 10^{-2}$~M$_\odot$/yr at $z=20$).  
This can be explained by considering 
the evolving virial properties of the halos with redshift and the physics of molecular 
hydrogen formation at low densities.  Our result implies that the mass distribution of Population 
III stars (as inferred from their accretion rates) may be broader than 
previously thought, and may evolve with redshift.   Finally, we observe that our collapsing 
protostellar cloud cores do not fragment, consistent with previous results, 
which suggests that Population III
stars which form in halos of mass $10^5 - 10^6$~M$_\odot$ always form in isolation.
\end{abstract}


\section{Introduction}\label{Intro}

The nature of the first generation of stars, and their influence on later epochs
of structure formation, is a fundamental issue in modern cosmology.  A great
deal of theoretical progress has been made (see reviews by~\markcite{2004ARA&A..42...79B}{Bromm} \& {Larson} (2004)
and~\markcite{2005SSRv..116..625C}{Ciardi} \& {Ferrara} (2005)).  However, a major question remains:  What is the initial 
mass function of the first stars?  The answer to this question will help 
explain the contributions that Population III stars made towards 
reionization and chemical enrichment of the universe, and will also determine if
these stars form compact objects (which may be the progenitors of the supermassive
black holes found in the centers of most massive galaxies), and the properties
of those objects.

The nature of the Population III IMF has been debated for over half a century
~\markcite{1953Obs....73...77S, 1971Ap&SS..14..399E,1983ApJ...271..632P,
1983MNRAS.205..705S,1997ApJ...474....1T}({Schwarzschild} \& {Spitzer} 1953; {Ezer} \& {Cameron} 1971; {Palla}, {Salpeter}, \&  {Stahler} 1983; {Silk} 1983; {Tegmark} {et~al.} 1997).  It has been generally agreed
upon that the lack of efficient cooling mechanisms results in very massive stars, 
with an IMF that is extremely top-heavy compared to the galactic IMF.  This has
been supported by recent cosmological simulations of the formation of
Population III stars, which argue for massive objects
\markcite{ABN02,2002ApJ...564...23B,2004NewA....9..353B,2006astro.ph..6106Y}({Abel} {et~al.} 2002; {Bromm} {et~al.} 2002; {Bromm} \& {Loeb} 2004; {Yoshida} {et~al.} 2006).  
These simulations find
that the Jeans-unstable clumps which form in the cores of high-redshift halos are
on the order of a thousand solar masses out of which a single
primordial protostar condenses, and the subsonic contraction of the clump
(even at high densities) greatly reduces the chance of fragmentation into many low-mass
stars.

Though suggestive, these cosmological simulations lack the necessary physics and resolution
to follow
the collapse of the primordial cloud core to the point where accretion onto the star has been stopped
and the star's final mass has been determined.  The issue of accretion shutoff onto Pop 
III stars has been examined by \markcite{2003ApJ...589..677O}{Omukai} \& {Palla} (2003) and \markcite{TanMcKee2004}{Tan} \& {McKee} (2004).
Omukai \& Palla perform 1D, spherically symmetric simulations of the shutoff of
accretion onto a primordial protostar.  They find that there is a critical mass
accretion rate, $\dot{m}_{crit} \simeq 4 \times 10^{-3}$~M$_\odot$/year, below which
stars can grow to essentially arbitrary masses (certainly $\geq 100$~M$_\odot$), 
limited only by the mass of available gas and the stellar lifetime.  Above this
critical accretion rate, the maximum mass of the star decreases as $\dot{m}$ increases due 
to radiative feedback.
They find that for a more realistic, time-varying accretion rate taken from 
~\markcite{ABN02}{Abel} {et~al.} (2002) (hereafter referred to as ABN02), 
 the star can grow to masses of $\gg 100$~M$_\odot$.  
Tan \& McKee present a theoretical model of Population III star formation that combines
  models of the collapsing cloud core, evolution of the
primordial protostar and accretion disk.  They show that radiative feedback
processes are extremely important once the protostar reaches a mass of $\simeq 30$~M$_\odot$.
Their models use as the fiducial case an accretion rate onto the disk taken from the simulation
of ABN02.  As with the Omukai \& Palla calculation, accretion rates onto
the forming protostar are critical input parameters, if not the most critical parameter,
for determining the final mass of the Population III star.

How well have the accretion rates onto Population III protostars been constrained?
Recent work has made it clear that these rates must be determined primarily
from cosmological simulations.  \markcite{2002ApJ...564...23B}{Bromm} {et~al.} (2002) do not have
the mass and spatial resolution necessary to provide constraints.  ABN02, in one of the 
most highly resolved simulations published to date, simulated the formation of a single
Population III protostellar core that has a maximum accretion rate onto
the forming protostar of $\simeq 10^{-2}$~M$_\odot$/year.  
\markcite{2004NewA....9..353B}{Bromm} \& {Loeb} (2004) perform a high-resolution calculation of the formation of
a primordial protostar and find an accretion rate rate that starts at 
$\simeq 0.1$~M$_\odot$/year, but declines rapidly.  Finally,~\markcite{2006astro.ph..6106Y}{Yoshida} {et~al.} (2006)
present a calculation of the formation of a primordial protostellar cloud that takes
into account opacity effects due to molecular hydrogen and also implements 
a novel particle-splitting scheme.  This allows them to accurately follow the
evolution of a primordial gas cloud which collapses to a density of 
$n_H \simeq 10^{16}$~cm$^{-3}$.  As in ABN02, Yoshida et al. do not observe 
fragmentation of the collapsing cloud core, and 
observe gas accretion rates that peak at $10^{-1}$~M$_\odot$/yr
and average around $10^{-3}-10^{-2}$~M$_\odot$/yr.

All of the published work thus far which studies
Population III star formation in a cosmological context and at extremely high resolution
has been an examination of
a single object, and as such cannot answer fundamental questions, such as: What is
the variance in accretion rates onto primordial protostars?  Is there some 
systematic dependence
upon redshift or environment?  and, finally, Is the lack of fragmentation in the 
collapsing halo core universal?  Some previous work
has studied large numbers of primordial halos~\markcite{2001ApJ...548..509M,
2003MNRAS.338..273M,2003ApJ...592..645Y}({Machacek}, {Bryan}, \&  {Abel} 2001, 2003; {Yoshida} {et~al.} 2003).  However, these studies lacked the 
resolution to study properties of the halo cores in detail.

In this work we will attempt to address these questions by performing a series of
extremely high dynamical range adaptive mesh refinement (AMR) simulations of
the formation of a Population III protostar in a $\Lambda$CDM cosmology.  Our mass and
spatial resolution is comparable to ABN02 and~\markcite{2006astro.ph..6106Y}{Yoshida} {et~al.} (2006).  This is the
first time that a significant number of such high-resolution calculations have been
sampled, allowing us to investigate the robustness of previous results and address
questions that they were unable to examine.  Our 
sample includes halos which cool and exhibit runaway collapse at their centers
over the redshift range $32 > z > 19$.  The halo virial masses at collapse range from 
$1.5-7 \times 10^5$~M$_\odot$ and show no systematic trend with redshift.  We explore the
environmental effects that lead to these variations.  We
discover that there is a trend towards increasing accretion rates in halos with decreasing
collapse redshifts (later collapse times), which is caused by evolving halo virial properties
and the physics of molecular hydrogen formation.
Peak accretion rates vary between
$10^{-4}$ and $10^{-2}$~M$_\odot$/year and have similar time
dependence.  Finally, we observe that our collapsing protostellar cloud cores do not
fragment, consistent with previous results, which suggests that Population III
stars which form in halos of mass $10^5 - 10^6$~M$_\odot$ always form in isolation.

The simulations discussed in this paper study Population III protostars
which form in halos of mass $10^5-10^6$~M$_\odot$ and in the redshift 
range $z \simeq 20-30$.  These halos are far more common than the extremely
 rare peaks discussed by~\markcite{2005MNRAS.363..393R}{Reed} {et~al.} (2005) and
~\markcite{2005MNRAS.363..379G}{Gao} {et~al.} (2005), and thus represent a more ``typical'' population of
primordial stars.  Additionally, in this paper we do not concern ourselves
with halos with virial temperatures above $10^4$ K, where cooling by atomic
hydrogen could be important~\markcite{2002ApJ...569..558O,2003ApJ...596...34B}({Oh} \& {Haiman} 2002; {Bromm} \& {Loeb} 2003).
Investigation of this mode of Population III star formation, though potentially
very important, is deferred to a later paper.

This is the first in a series of papers presenting results of high dynamical
range AMR cosmological simulations of the formation of Population
III stars in a $\Lambda$CDM cosmological context.  As stated previously,
this paper discusses issues relating to accretion rates onto evolving primordial
protostellar cores.  The second paper in this series (O'Shea, Norman \& Li 2006, in 
preparation) presents an investigation of the evolution of angular momentum in a collapsing
primordial halo core.  The third paper in this series (O'Shea \& Norman 2006, in preparation)
presents an examination of the formation of Population III protostellar halos in the
presence of a molecular hydrogen photodissociating UV background.  Later papers 
will present more detailed investigations of the evolution of the collapsing protostellar
cores and the formation of a primordial protostar embedded in a thick accretion disk.

The organization of this paper is as follows:
In Section~\ref{Methodology}, we provide a summary of Enzo, the code used to perform
the calculations in this paper and the second and third papers in this series, 
and of the calculation setup.  Sections~\ref{repstar} and ~\ref{compare-real} summarize
our results:  In Section~\ref{repstar}, we present the detailed evolution
of a single, representative Population III protostellar cloud, and compare it to 
previous work.  In Section~\ref{compare-real}, we present results of a series of 
calculations of Population III protostellar cloud formation with varied large scale 
structure and simulation volumes, and show
some systematic trends in the properties of these clouds.  In Section~\ref{discuss}
we discuss the results presented in this work, and in Section~\ref{summary} we
present a summary of the main results and conclusions.

\section{Methodology}\label{Methodology}

\subsection{The Enzo code}\label{enzocode}

`Enzo'\footnote{http://lca.ucsd.edu/codes/currentcodes/enzo} is a publicly available, extensively tested 
adaptive mesh refinement
cosmology code developed by Greg Bryan and others \markcite{bryan97,bryan99,norman99,oshea04,
2005ApJS..160....1O}({Bryan} \& {Norman} 1997a, 1997b; {Norman} \& {Bryan} 1999; {O'Shea} {et~al.} 2004, 2005).
The specifics of the Enzo code are described in detail in these papers (and references therein),
but we present a brief description here for clarity.

The Enzo code couples an N-body particle-mesh (PM) solver \markcite{Efstathiou85, Hockney88}({Efstathiou} {et~al.} 1985; {Hockney} \& {Eastwood} 1988) 
used to follow the evolution of a collisionless dark
matter component with an Eulerian AMR method for ideal gas dynamics by \markcite{Berger89}{Berger} \& {Colella} (1989), 
which allows high dynamic range in gravitational physics and hydrodynamics in an 
expanding universe.  This AMR method (referred to as \textit{structured} AMR) utilizes
an adaptive hierarchy of grid patches at varying levels of resolution.  Each
rectangular grid patch (referred to as a ``grid'') covers some region of space in its
\textit{parent grid} which requires higher resolution, and can itself become the 
parent grid to an even more highly resolved \textit{child grid}.  Enzo's implementation
of structure AMR places no fundamental restrictions on the number of grids at a 
given level of refinement, or on the number of levels of refinement.  However, owing 
to limited computational resources it is practical to institute a maximum level of 
refinement, $\ell_{max}$.  Additionally, the Enzo AMR implementation allows arbitrary 
integer ratios of parent
and child grid resolution, though in general for cosmological simulations (including the 
work described in this paper) a refinement ratio of 2 is used.

Since the addition of more highly refined grids is adaptive, the conditions for refinement 
must be specified.  In Enzo, the criteria for refinement can be set by the user to be
a combination of any or all of the following:  baryon or dark matter overdensity
threshold, minimum resolution of the local Jeans length, local density gradients,
local pressure gradients, local energy gradients, shocks, and cooling time.
A cell reaching
any or all of the user-specified criteria will then be flagged for refinement.  Once all 
cells of a given level have been flagged, rectangular solid boundaries are determined which 
minimally 
encompass them.  A refined grid patch is then introduced within each such bounding 
volume, and the results are interpolated to a higher level of resolution.

In Enzo, resolution of the equations being solved is adaptive in time as well as in
space.  The timestep in Enzo is satisfied on a level-by-level basis by finding the
largest timestep such that the Courant condition (and an analogous condition for 
the dark matter particles) is satisfied by every cell on that level.  All cells
on a given level are advanced using the same timestep.  Once a level $L$ has been
advanced in time $\Delta t_L$, all grids at level $L+1$ are 
advanced, using the same criteria for timestep calculations described above, until they
reach the same physical time as the grids at level $L$.  At this point grids at level
$L+1$ exchange baryon flux information with their parent grids, providing a more 
accurate solution on level $L$.  Cells at level $L+1$ are then examined to see 
if they should be refined or de-refined, and the entire grid hierarchy is rebuilt 
at that level (including all more highly refined levels).  The timestepping and 
hierarchy rebuilding process is repeated recursively on every level to the 
maximum existing grid level in the simulation.

Two different hydrodynamic methods are implemented in Enzo: the piecewise parabolic
method (PPM) \markcite{Woodward84}({Woodward} \& {Colella} 1984), which was extended to cosmology by 
\markcite{Bryan95}{Bryan} {et~al.} (1995), and the hydrodynamic method used in the ZEUS magnetohydrodynamics code
\markcite{stone92a,stone92b}({Stone} \& {Norman} 1992a, 1992b).  We direct the interested reader to the papers describing 
both of these methods for more information, and note that PPM is the preferred choice
of hydro method since it is higher-order-accurate and is based on a technique that 
does not require artificial viscosity, which smoothes shocks and can smear out 
features in the hydrodynamic flow.

The chemical and cooling properties of primordial (metal-free) gas are followed 
using the method of \markcite{abel97}{Abel} {et~al.} (1997) and \markcite{anninos97}{Anninos} {et~al.} (1997), supplemented with reactions for 
the 3-body
formation of $H_2$ documented in ABN02.  
This method follows the non-equilibrium evolution of a 
gas of primordial composition with 9 total species:  
$H$, $H^+$, $He$, $He^+$, $He^{++}$, $H^-$, $H_2^+$, $H_2$, and $e^-$.  The code 
also calculates 
radiative heating and cooling following atomic line excitation, recombination,
collisional excitation, free-free transitions, molecular line excitations, and Compton
scattering of the cosmic microwave background, as well as any of
approximately a dozen different models for a metagalactic UV background that heat
the gas via photoionization and photodissociation.  We model the cooling processes
detailed in ~\markcite{abel97}{Abel} {et~al.} (1997), but with the molecular hydrogen cooling function
of~\markcite{1998A&A...335..403G}{Galli} \& {Palla} (1998). 
The multispecies rate equations are solved out of
equilibrium to properly model situations where, e.g., the cooling time of the gas
is much shorter than the hydrogen recombination time.  
A total of 9 kinetic equations are solved, including 29 kinetic and radiative 
processes, for the 9 species mentioned above.  
The chemical reaction equation network is technically challenging to solve due to 
the huge range of reaction time scales involved -- the characteristic creation
and destruction time scales of the various species and reactions can differ by 
many orders of magnitude.  As a result, the set of rate equations is extremely 
stiff, and an explicit scheme for integration of the rate equations can be 
costly if small enough timestep are taken to keep the network
stable.  This makes an implicit scheme preferable for such a set of 
equations, and Enzo solves the rate equations using a method based on a backwards 
differencing formula (BDF) in order to provide a stable and accurate solution.
 
It is important to note the regime in which this chemistry model is valid.  According to 
\markcite{abel97}{Abel} {et~al.} (1997) and \markcite{anninos97}{Anninos} {et~al.} (1997), the reaction network is valid for temperatures
between $10^0 - 10^8$ K.  The original model discussed in these two references is only
valid up to n$_H \sim 10^4$~cm$^{-3}$.  However, addition of the 3-body H$_2$ formation
process allows correct modeling of the chemistry of the gas up until the point 
where collisionally-induced emission from molecular hydrogen becomes an important
cooling process, which occurs at n$_H \sim 10^{14}$~cm$^{-3}$.  We do not include 
heating by molecular hydrogen formation,
which should take place at densities above $10^8$~$cm^{-3}$, and may affect 
temperature evolution at these high densities.  A further concern is that
the optically thin approximation for radiative cooling breaks down, which
begins at n$_H \sim 10^{10} - 10^{12}$~cm$^{-3}$.  Beyond this point, 
modifications to the cooling function that take into account the non-negligible
opacity of the gas to line radiation from molecular hydrogen must be made, as 
discussed by \markcite{ripamonti04}{Ripamonti} \& {Abel} (2004).  Even with these modifications, a more correct 
description of the cooling of gas of primordial composition at high densities will 
require some form of radiation transport, which will greatly 
increase the cost of the simulations.

\subsection{Simulation setup}\label{simsetup}

The simulations discussed in this paper are set up as follows.  We initialize all 
calculations at $z=99$ assuming a ``concordance'' cosmological model:  $\Omega_m = 0.3$, 
$\Omega_b = 0.04$, $\Omega_{CDM} = 0.26$, $\Omega_\Lambda = 0.7$, $h=0.7$ (in units of 100 km/s/Mpc), 
$\sigma_8 = 0.9$, and using an \markcite{eishu99}{Eisenstein} \& {Hu} (1999) power spectrum
with a spectral index of $n = 1$.  Twelve simulations are 
generated using a separate random seed for each, meaning that the large-scale structure
that forms in each of the simulation volumes is statistically independent of the others.
These simulations, detailed in Table~\ref{table.siminfo}, 
are divided into sets of four simulations in three different
box sizes: $0.3, 0.45$, and~$0.6$~h$^{-1}$~Mpc (comoving). 
The most massive halo to form in each 
simulation at $z=15$ (typically with a mass of $\sim 10^6$~M$_\odot$) is found using a dark 
matter-only calculation
with $128^3$ particles on a $128^3$ root grid with a maximum of 4 levels of adaptive mesh, refining on 
a dark matter overdensity criterion of $8.0$.  The initial conditions are then regenerated
with both dark matter and baryons for each of the simulation volumes 
such that the Lagrangian volume in which the halo formed
is now resolved at much higher spatial and mass resolution.  These simulations
have a $128^3$ 
root grid and three levels of static nested grids, each with twice
the spatial resolution and eight times the mass resolution of the previous grid.  This gives
an overall effective grid size of $1024^3$
in the region where the most massive halo will form.
The highest resolution grid in each simulation is $256^3$ grid cells, and corresponds
to a volume $(75, 112.5, 150$)~h$^{-1}$ comoving kpc on a side for the 
(0.3, 0.45, 0.6)~h$^{-1}$ Mpc box.  The dark matter particles in the highest
resolution grid are (1.81, 6.13, 14.5)~h$^{-1}$~M$_\odot$ and the spatial resolution
of cells on these grids are (293, 439, 586)~h$^{-1}$ parsecs (comoving).  Though the simulations
have a range of initial spatial resolutions and dark matter masses, we find that the
final simulation results are converged -- the spatial and mass resolution of the 
0.3~h$^{-1}$ Mpc volume simulations can be degraded to that of the 0.6~h$^{-1}$ Mpc
without significantly changing the results.

The simulations with nested grid initial conditions are then 
started at $z=99$ and allowed to evolve until the collapse
of the gas within the center of the most massive halo, which occurs at a range of
redshifts (as shown in Section~\ref{compare-real}, and summarized in 
Tables~\ref{table.meaninfo-1} and~\ref{table.meaninfo-2}).  The equations of hydrodynamics
are solved using the PPM method with a dual energy formulation (the results 
are essentially identical when the
ZEUS hydrodynamic method is used).  The nonequilibrium chemical
evolution and optically thin radiative cooling of the primordial gas is 
modeled as described in Section~\ref{enzocode}, following 9 
separate species including molecular hydrogen (but excluding deuterium).  Adaptive
mesh refinement is enabled within the high resolution region
 such that cells are refined by factors of two along each 
axis, to a maximum of 22 total levels of refinement.  This corresponds to a 
maximum resolution of (115, 173, 230)~h$^{-1}$ astronomical units (comoving)
at the finest level of resolution, with an overall spatial dynamical range of
$5.37 \times 10^8$.  To avoid effects due to the finite size of the dark matter
particles, the dark matter density is smoothed on a comoving scale of $\sim 0.5$~pc
(a proper scale of $\sim 0.025$ pc at $z \sim 20$).
This is reasonable because at that radius in all of our calculations the gravitational
potential is completely dominated by the baryons.

Grid cells are adaptively refined based upon several criteria:
baryon and dark matter overdensities in cells of 4.0 and 8.0, respectively, checks 
to ensure that the pressure jump and/or energy ratios between adjoining
cells never exceeds 5.0, that the cooling time in a given cell is always longer
than the sound crossing time of that cell, and that the Jeans length is always
resolved by at least 16 cells.  This guarantees that the Truelove
criterion~\markcite{truelove97}({Truelove} {et~al.} 1997), which is an empirical result showing that in order 
to avoid artificial gravitational fragmentation
in numerical simulations the Jeans length must be resolved by at least 4 grid cells,
 is always maintained by a significant margin.
Simulations which force the Jeans length to be resolved by a minimum of 4 and 64
cells produce results which are essentially identical to those presented here.
\clearpage
\begin{deluxetable}{cccccccccc}
\tablecolumns{10}
\tablewidth{0pt}
\tablecaption{Simulation Parameters}
\tablehead{\colhead{Run} & \colhead{$\L_{box}$} & \colhead{ $\ell_{min}$ } & \colhead{$m_{\rm DM}$} 
& \colhead{$m_{\rm gas,init}$} & \colhead{$m_{\rm gas,max}$} & \colhead{$\sigma_\rho$} & 
\colhead{$\delta_{20}$} & \colhead{$\delta_{40}$} & \colhead{$\sigma_{halo}$} }
\startdata
L0\_30A & 0.3  &  115.3  & 2.60 & 0.40 & $5.43\times 10^{-4}$ & 0.1964 & 1.873 & 1.153 & 2.557 \\
L0\_30B & 0.3  &  115.3  & 2.60 & 0.40 & $5.43\times 10^{-4}$ & 0.1952 & 2.176 & 1.322 & 2.858 \\
L0\_30C & 0.3  &  115.3  & 2.60 & 0.40 & $5.43\times 10^{-4}$ & 0.1995 & 1.981 & 1.319 & 2.778 \\
L0\_30D & 0.3  &  115.3  & 2.60 & 0.40 & $5.43\times 10^{-4}$ & 0.1922 & 2.727 & 1.702 & 3.022 \\
\tableline \\
L0\_45A & 0.45 &  173.0  & 8.76 & 1.35 & $1.83\times 10^{-3}$ & 0.1879 & 2.669 & 1.873 & 3.615 \\
L0\_45B & 0.45 &  173.0  & 8.76 & 1.35 & $1.83\times 10^{-3}$ & 0.1860 & 2.726 & 1.872 & 3.343 \\
L0\_45C & 0.45 &  173.0  & 8.76 & 1.35 & $1.83\times 10^{-3}$ & 0.1906 & 2.375 & 1.563 & 3.293 \\
L0\_45D & 0.45 &  173.0  & 8.76 & 1.35 & $1.83\times 10^{-3}$ & 0.1861 & 1.959 & 1.191 & 3.827 \\
\tableline \\
L0\_60A & 0.6  &  230.6  & 20.8 & 3.20 & $4.34\times 10^{-3}$ & 0.1815 & 2.458 & 1.790 & 3.154 \\
L0\_60B & 0.6  &  230.6  & 20.8 & 3.20 & $4.34\times 10^{-3}$ & 0.1802 & 3.049 & 1.956 & 3.340 \\
L0\_60C & 0.6  &  230.6  & 20.8 & 3.20 & $4.34\times 10^{-3}$ & 0.1815 & 2.840 & 2.121 & 3.466 \\
L0\_60D & 0.6  &  230.6  & 20.8 & 3.20 & $4.34\times 10^{-3}$ & 0.1959 & 2.700 & 1.814 & 4.147 \\
\enddata
\tablecomments{ 
$L_{box}$ is the size of the simulation volume in comoving $h^{-1}$~Mpc.
$\ell_{min}$ is the minimum comoving spatial resolution element in comoving astronomical units (AU).
$m_{\rm DM}$ is the dark matter particle mass within the highest-resolution static nested grid in units of M$_\odot$.
$m_{\rm gas,init}$ is the mean baryon mass resolution within the highest-resolution static nested grid in
 units of $M_\odot$, at the beginning of the simulation.
$m_{\rm gas,max}$ is the mean baryon mass resolution within the highest level grid cell at the end of the simulation in
units of M$_\odot$.
$\sigma_\rho$ is the dispersion in overdensities of the baryon gas within the highest-resolution static nested
grid at the beginning of the simulation.
$\delta_{20}$ and $\delta_{40}$ are the mean overdensity of all gas within 20 and 40 virial radii of the halo
whose core is collapsing, at the epoch of collapse (calculated as described in the text), 
in units of $\Omega_b \rho_c$.
$\sigma_{halo}$ is the statistical probability of the halo in which the primordial protostellar cloud forms, 
at the redshift of collapse, assuming Gaussian statistics and the same Eisenstein \& Hu CDM power 
spectrum as is used to initialize our calculations \markcite{eishu99}({Eisenstein} \& {Hu} 1999).
}
\label{table.siminfo}
\end{deluxetable}

\begin{deluxetable}{cccccccccc}
\tablecolumns{10}
\tablewidth{0pt}
\tablecaption{Halo Properties}
\tablehead{\colhead{Run} & \colhead{$z_{coll}$} & \colhead{$M_{vir}$} &  \colhead{$R_{vir}$} 
& \colhead{$T_{vir}$} & \colhead{ $M_{b}$ } & \colhead{$f_{bar}$} 
 & \colhead{$\lambda_{dm}$}  & \colhead{$\lambda_{gas}$}  & \colhead{$\theta$} }
\startdata
L0\_30A & 19.28 & $4.18 \times 10^5$ & 118.1 & 1120.0 & $4.59 \times 10^4$ & 0.823 & 0.050 & 0.022 & 55.1 \\
L0\_30B & 22.19 & $2.92 \times 10^5$ & 91.7  & 1009.6 & $2.91 \times 10^4$ & 0.753 & 0.066 & 0.029 & 3.3  \\ 
L0\_30C & 20.31 & $6.92 \times 10^5$ & 132.9 & 1646.8 & $7.56 \times 10^4$ & 0.819 & 0.053 & 0.034 & 17.3 \\ 
L0\_30D & 24.74 & $1.36 \times 10^5$ & 64.0  & 671.4  & $1.29 \times 10^4$ & 0.711 & 0.046 & 0.014 & 12.7 \\ 
\tableline \\
L0\_45A & 28.70 & $2.41 \times 10^5$ & 67.4  & 1131.7 & $2.03 \times 10^4$ & 0.631 & 0.108 & 0.056 & 8.4  \\ 
L0\_45B & 26.46 & $2.42 \times 10^5$ & 72.7  & 1052.8 & $2.28 \times 10^4$ & 0.706 & 0.019 & 0.019 & 34.0 \\ 
L0\_45C & 24.74 & $5.21 \times 10^5$ & 100.1 & 1645.6 & $5.21 \times 10^4$ & 0.750 & 0.054 & 0.015 & 97.4 \\
L0\_45D & 28.67 & $5.89 \times 10^5$ & 90.5  & 2060.0 & $5.71 \times 10^4$ & 0.727 & 0.059 & 0.071 & 11.4 \\
\tableline \\
L0\_60A & 24.10 & $3.96 \times 10^5$ & 93.8  & 1338.0 & $4.29 \times 10^4$ & 0.813 & 0.052 & 0.027 & 26.0 \\ 
L0\_60B & 25.64 & $3.82 \times 10^5$ & 87.3  & 1385.6 & $3.72 \times 10^4$ & 0.730 & 0.047 & 0.056 & 11.0 \\ 
L0\_60C & 28.13 & $1.68 \times 10^5$ & 60.7  & 876.9  & $1.29 \times 10^4$ & 0.575 & 0.027 & 0.016 & 40.0 \\ 
L0\_60D & 32.70 & $2.85 \times 10^5$ & 62.5  & 1440.9 & $2.06 \times 10^4$ & 0.542 & 0.049 & 0.033 & 10.0 \\ 
\enddata
\tablecomments{ Mean halo properties at the collapse redshift.
$z_{coll}$ is the collapse redshift of the halo, defined as the redshift at which the central baryon
density of the halo reaches $n_H \simeq 10^{10}$ cm$^{-3}$.
$M_{vir}$, $R_{vir}$, and $T_{vir}$ are the halo virial mass, radius, and temperature at that epoch, respectively.
$M_{b}$ is the total baryon mass within the virial radius at that epoch, and $f_{bar}$ is the baryon mass fraction 
(in units of $\Omega_b/\Omega_m$). 
$\lambda_{dm}$ and $\lambda_{gas}$ are the halo dark matter and gas spin parameters, respectively, and
$\theta$ is the angle of separation (in degrees) between the bulk dark matter and gas angular momentum 
vectors.  $M_{vir}$ and $M_b$ are in units of $M_\odot$, $R_{vir}$ is in units of proper parsecs,
and $T_{vir}$ is in units of degrees Kelvin.
}
\label{table.meaninfo-1}
\end{deluxetable}

\begin{deluxetable}{cccccccc}
\tablecolumns{8}
\tablewidth{0pt}
\tablecaption{Core Properties}
\tablehead{\colhead{Run} & 
\colhead{$\dot{m}_{max}$}  & \colhead{$<\dot{m}>_m$} & \colhead{$<T_{halo}>_m$} & \colhead{$T_{core}$} & 
\colhead{$<f_{H2}>_m$} & \colhead{$f_{H2,core}$} & \colhead{$M_{be}$} }
\startdata
L0\_30A & $1.45\times 10^{-2}$ & $2.52 \times 10^{-3}$ & 745.8 & 438.3 & $9.31 \times 10^{-4}$ & $9.12 \times 10^{-4}$ & 1159.4 \\
L0\_30B & $1.74\times 10^{-2}$ & $6.69 \times 10^{-3}$ & 645.2 & 468.6 & $8.00 \times 10^{-4}$ & $1.05 \times 10^{-3}$ & 1768.6 \\ 
L0\_30C & $3.02\times 10^{-2}$ & $1.59 \times 10^{-2}$ & 808.1 & 501.7 & $7.93 \times 10^{-4}$ & $1.01 \times 10^{-3}$ & 1797.7 \\ 
L0\_30D & $6.47\times 10^{-3}$ & $7.08 \times 10^{-4}$ & 512.9 & 310.7 & $1.14 \times 10^{-3}$ & $7.58 \times 10^{-4}$ & 814.8  \\ 
\tableline \\
L0\_45A & $6.79\times 10^{-4}$ & $1.52 \times 10^{-4}$ & 858.4 & 276.3 & $1.78 \times 10^{-3}$ & $1.73 \times 10^{-3}$ & 1120.2 \\ 
L0\_45B & $7.64\times 10^{-4}$ & $2.92 \times 10^{-4}$ & 698.0 & 295.3 & $9.97 \times 10^{-4}$ & $9.97 \times 10^{-4}$ & 1077.3 \\ 
L0\_45C & $5.92\times 10^{-3}$ & $2.78 \times 10^{-3}$ & 897.8 & 502.7 & $1.05 \times 10^{-3}$ & $7.09 \times 10^{-4}$ & 1141.0 \\
L0\_45D & $7.52\times 10^{-3}$ & $2.25 \times 10^{-3}$ & 878.0 & 354.2 & $1.02 \times 10^{-3}$ & $1.04 \times 10^{-3}$ & 856.6  \\
\tableline \\
L0\_60A & $2.45\times 10^{-3}$ & $9.05 \times 10^{-4}$ & 808.8 & 431.1 & $1.45 \times 10^{-3}$ & $8.17 \times 10^{-4}$ & 1750.8 \\ 
L0\_60B & $9.74\times 10^{-4}$ & $1.52 \times 10^{-4}$ & 803.3 & 300.1 & $1.60 \times 10^{-3}$ & $1.14 \times 10^{-3}$ & 1301.2 \\ 
L0\_60C & $4.72\times 10^{-4}$ & $2.66 \times 10^{-4}$ & 753.7 & 420.1 & $1.41 \times 10^{-3}$ & $1.04 \times 10^{-3}$ & 1395.9 \\ 
L0\_60D & $3.77\times 10^{-4}$ & $1.46 \times 10^{-4}$ & 917.2 & 249.5 & $1.88 \times 10^{-3}$ & $2.26 \times 10^{-3}$ & 928.5  \\ 
\enddata
\tablecomments{ Mean core properties at the collapse redshift.
$\dot{m}_{max}$ is the maximum accretion rate (in units of M$_\odot$/year) of gas onto the protostellar core, assuming no
feedback.
$<\dot{m}>_m$ is the mean, mass-weighted accretion rate (in units of M$_\odot$/year) of the inner $10^3$~M$_\odot$ of gas onto 
the protostellar core, assuming no feedback.
$<T_{halo}>_m$ is the mass-weighted mean temperature of all gas within the virial radius.
$T_{core}$ is the ``core'' gas temperature, defined as the temperature at $r=0.1$ pc (proper) at the redshift of collapse.
$<f_{H2}>_m$ is the mass-weighted molecular hydrogen fraction of all gas within the virial radius.
$f_{H2,core}$ is the ``core'' $H_2$ fraction, defined as the molecular hydrogen fraction at $r=0.1$ pc (proper) at the redshift
of collapse.
$M_{be}$ is the Bonnor-Ebert mass of the halo core at the epoch at which the central core baryon number 
density is approximately $10^4$~cm$^{-3}$.
}
\label{table.meaninfo-2}
\end{deluxetable}


\clearpage

\section{Evolution of a representative protostellar core}\label{repstar}

In this section we describe in detail the collapse of a single primordial 
protostellar core out of the ensemble discussed in Section~\ref{compare-real}.  This
simulation, L0\_30A,  was selected out of the four simulations performed in a
 $0.3$~h$^{-1}$ Mpc comoving volume.  The results described here are qualitatively
similar for all of the calculations described in Section~\ref{compare-real}, 
though there is some scatter in the exact evolution of each halo due to differences
in large scale structure and the detailed merger history of the halo.  However, since
the collapse is essentially controlled by the chemistry of molecular 
hydrogen formation, the result is general.

\begin{figure}
\begin{center}
\includegraphics[width=0.9\textwidth]{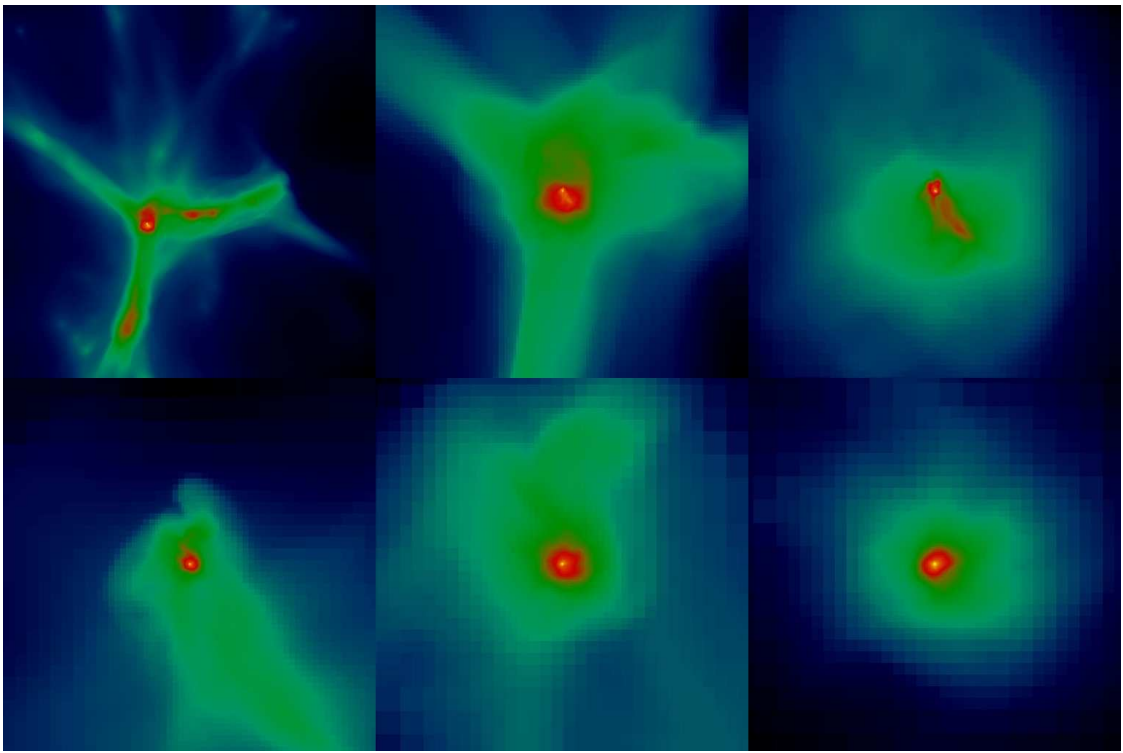}
\end{center}
\caption{
Zoom on projected mass-weighted baryon density by factors of four for a representative 
Population III protostar formation calculation at the last simulation output.  
At this redshift ($z=19.28$), 
the maximum density was $\sim 10^{12}$~cm$^{-3}$ with a cosmic mean density of 
$\simeq 0.003$~cm$^{-3}$, for an overall density increase
of 15 orders of magnitude.  Top left: view is 1320 pc across.  Top center: 330 pc. 
Top right:  82.5 pc.  Bottom left: 20.6 pc.  Bottom center: 5.2 pc.  Bottom right: 1.29
pc.  Note that all sizes are in proper parsecs at $z=19.28$.  In all panels yellow
represents high densities and blue represents low density, with the color table relative
in each frame.
}
\label{fig.rep.proj-dens}
\end{figure}

\begin{figure}
\begin{center}
\includegraphics[width=0.9\textwidth]{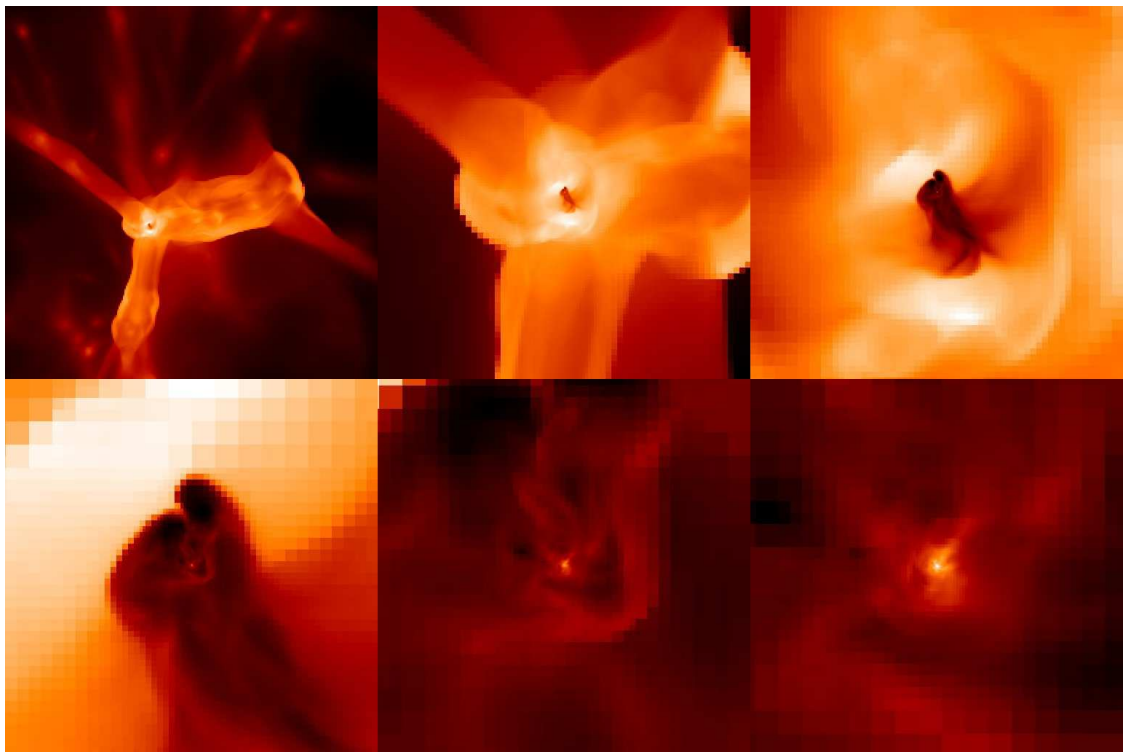}
\end{center}
\caption{
Zoom on projected mass-weighted baryon temperature by factors of four in a representative
Population III protostar formation calculation at the last simulation output.  The collapse
redshift is $z=19.28$ and the simulation and spatial sizes of each panel are the same as in 
Figure~\ref{fig.rep.proj-dens}.  In all panels white represents high temperatures and dark
colors represent low temperatures.  The color table is relative in each frame.
}
\label{fig.rep.proj-temp}
\end{figure}

\begin{figure}
\begin{center}
\includegraphics[width=0.9\textwidth]{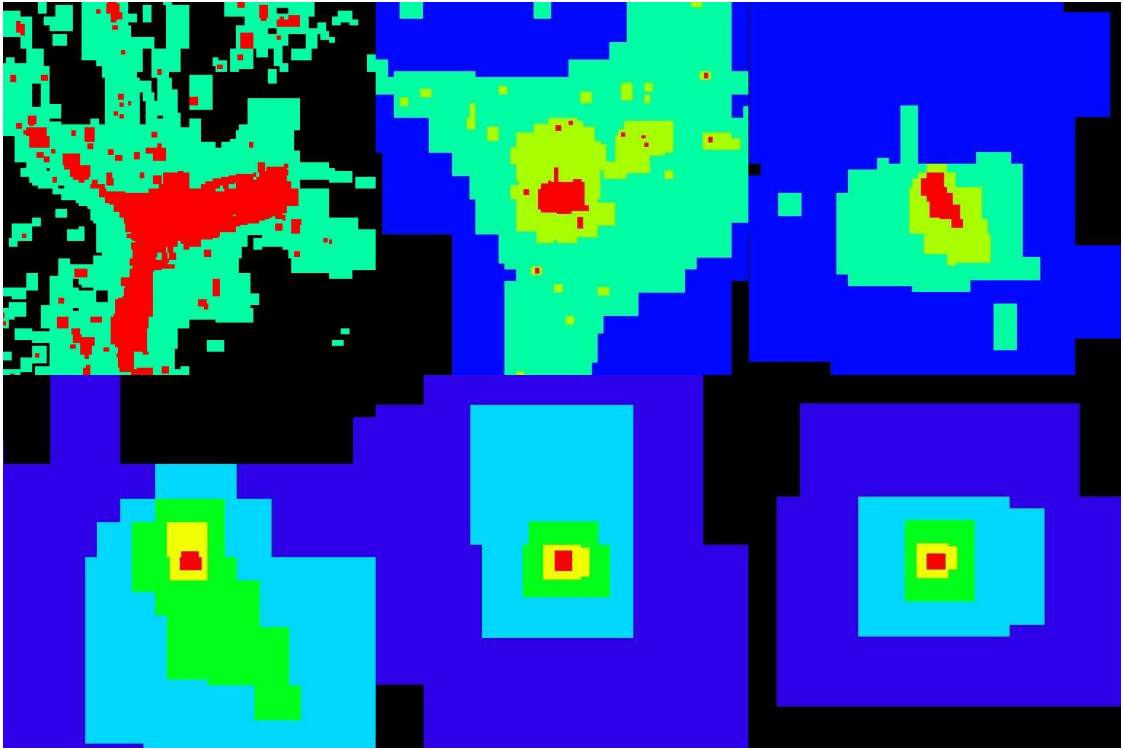}
\end{center}
\caption{
Zoom on projected maximum level in a representative Population III protostar 
formation calculation at the last simulation output.  The spatial scale for each panel
and simulation are the same as in Figure~\ref{fig.rep.proj-dens}.  The maximum projected
level in each panel is as follows.  Top left:  Level 6.  Top center:  Level 8.  Top right:
Level 10.  Bottom left:  Level 12.  Bottom center:  Level 14.  Bottom right:  Level 16. 
In each panel the highest level grid is represented is in red, second highest in yellow, 
and so on.  The highest level of resolution at this time is $L=22$.
}
\label{fig.rep.proj-level}
\end{figure}

Figures~\ref{fig.rep.proj-dens},~\ref{fig.rep.proj-temp},~and~\ref{fig.rep.proj-level}
zoom in on the central gas core in each halo at the redshift of collapse (defined as
the redshift where the baryon density at the halo center reaches $\simeq 10^{10}$~cm$^{-3}$)
by factors of four, showing projections of log baryon density, log baryon temperature, and
maximum refinement level, respectively.  The largest-scale panel shows a projection of 
a volume of the universe
1320 proper parsecs across and deep, and zooms in by factors of four to 
approximately 1.3 pc across.  Each 
panel is centered on the collapsing protostar.  At large scales it is apparent
from Figure~\ref{fig.rep.proj-dens} that the halo in which the first star in the 
simulation volume forms is at the intersection of two
cosmological filaments, a distinctly asymmetrical situation.  
Examination of Figure~\ref{fig.rep.proj-temp} shows that the
filaments and majority of the volume of the halo are relatively hot ($\sim 1000$~
Kelvin), due primarily to accretion shocks formed by gas raining onto the filaments
and into the halo.  However, as we zoom in towards the center of the halo we can see that
the high-density gas is at a much lower temperature (a few hundred Kelvin) due to 
cooling by the significant quantity of molecular hydrogen that is formed in the halo.
The gas within the halo is not particularly spherical until scales of a few parsecs
are reached, where a somewhat warmer core of gas (T~$\sim 1000$~K) forms with an overall mass of
a few thousand solar masses, harboring a fully-molecular protostellar cloud core with a mass of
$\sim 1$~M$_\odot$.  The central core is generally spheroidal due to gas 
pressure and is not rotationally supported at any point up to the time the simulation is 
stopped.  Note that at later times, when the core has collapsed further, it is likely that a 
rotationally-supported accretion disk will form -- see~\markcite{TanMcKee2004}{Tan} \& {McKee} (2004) for more details.  
Figure~\ref{fig.rep.proj-level} 
shows how the adaptive mesh refinement is
used to resolve the cosmological structure by concentrating refinement only where it is 
needed.  This method is extremely effective at conserving computational resources - 
the level 16 grids,
which are the highest level of resolution shown in Figure~\ref{fig.rep.proj-level}, only 
encompass $\sim 2.5 \times 10^{-17}$ of the overall volume!
\clearpage

\begin{figure}
\begin{center}
\includegraphics[width=0.9\textwidth]{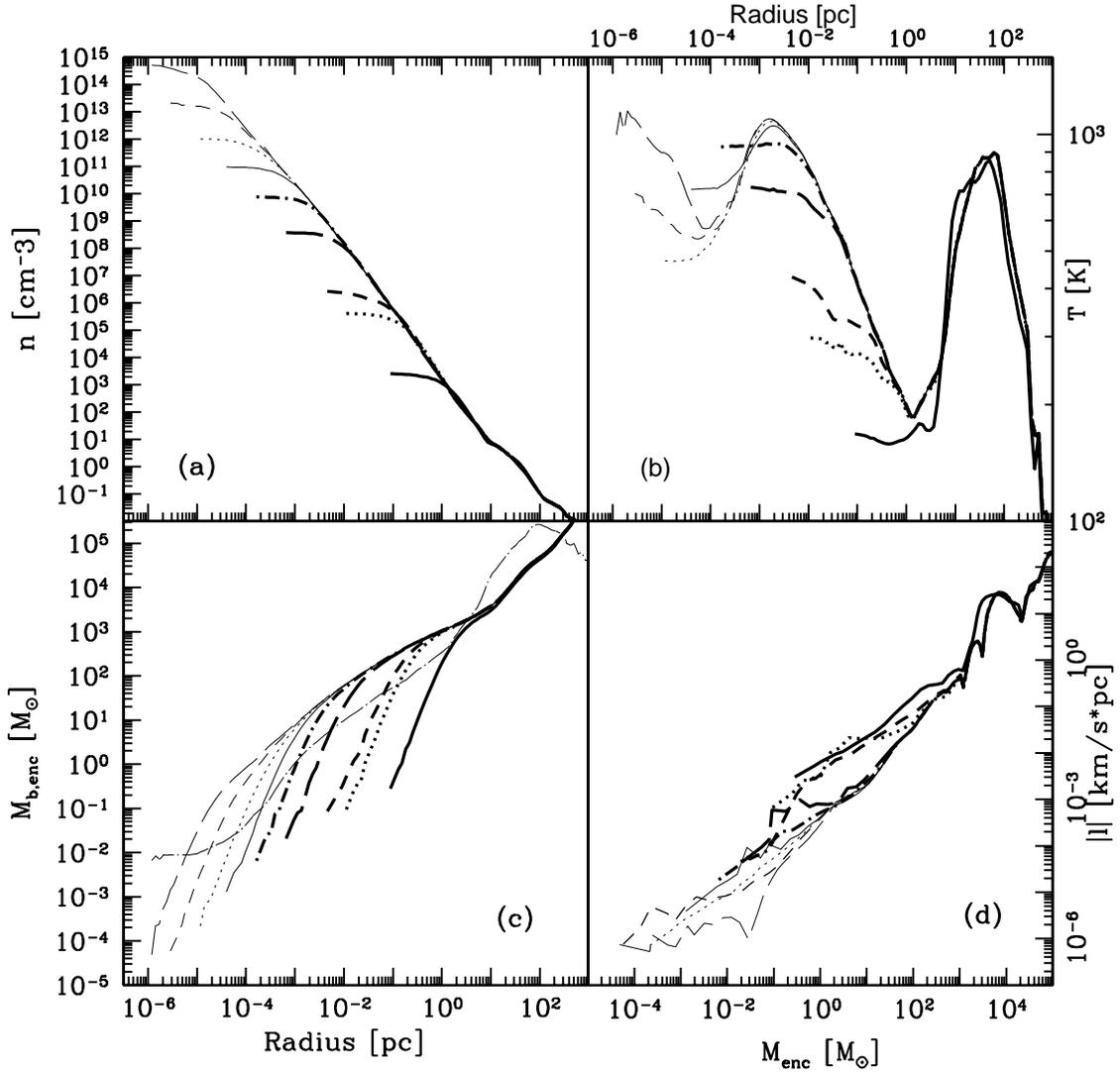}
\end{center}
\caption{
Evolution of spherically-averaged values for 
baryon number density (a), baryon temperature (b), 
and enclosed baryon mass (c) as a function of radius and
spherically-averaged baryon specific angular momentum as a function of enclosed mass 
(d) of a representative primordial protostar.
The thick solid black line in each panel corresponds to spherically averaged radial profile
of each quantity the onset of halo collapse, at
$z=18.05$ (approximately $2 \times 10^8$ years after the Big Bang).  Thick dotted
 line: the state of the halo $5.97 \times 10^6$ years after that.  Thick short-dashed line:
$1.89 \times 10^5$ years later.  Thick long-dashed line:  $1.21 \times 10^5$ years later. Thick
dot-dashed line: $6,810$ years later.  Thin solid line:  $782$ years later.  Thin dotted 
line: $151$ years later.  Thin short-dashed line: $58$ years later.  Thin long-dashed line: $14$ years
later, at a final output redshift of $z=17.673$.  The total time spanned by the lines in these
panels is $6.294 \times 10^6$ years.  The black dot-dashed line in the bottom left panel is
the Bonnor-Ebert critical mass calculated at the last timestep.}
\label{fig.rep.panel1}
\end{figure}

\begin{figure}
\begin{center}
\includegraphics[width=0.9\textwidth]{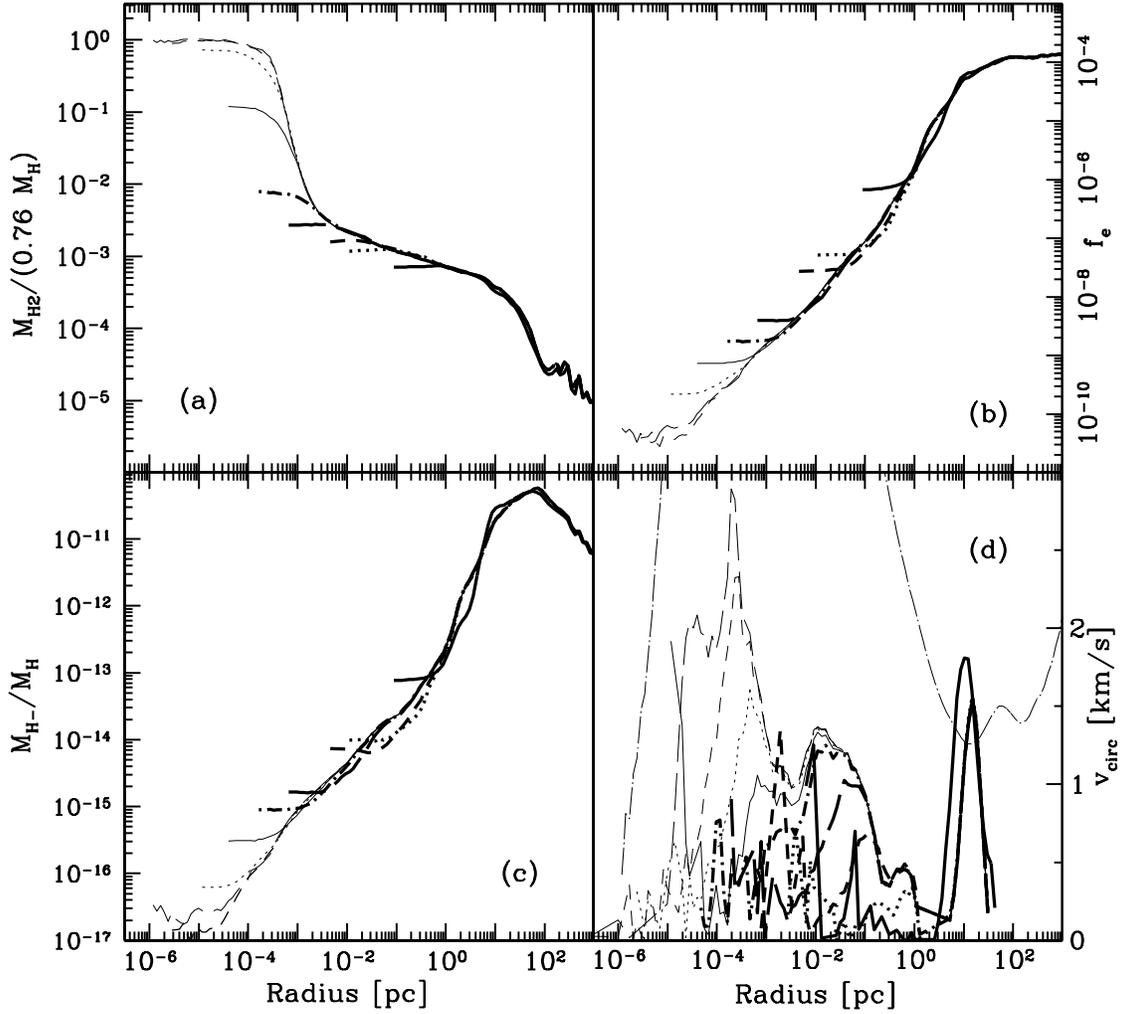}
\end{center}
\caption{
Evolution of spherically averaged radial profiles of 
molecular hydrogen fraction (a), electron fraction (b),
 $H^-$ fraction (c) and baryon circular velocity (d) as 
a function of radius of a representative primordial protostar.
The lines correspond to the same times as in Figure~\ref{fig.rep.panel1} and
are of the same simulation.   The black dot-dashed line in the plot of baryon circular 
velocity vs. time is the Newtonian circular velocity computed from the radius and
enclosed baryon mass at the last timestep.
}
\label{fig.rep.panel2}
\end{figure}

\begin{figure}
\begin{center}
\includegraphics[width=0.9\textwidth]{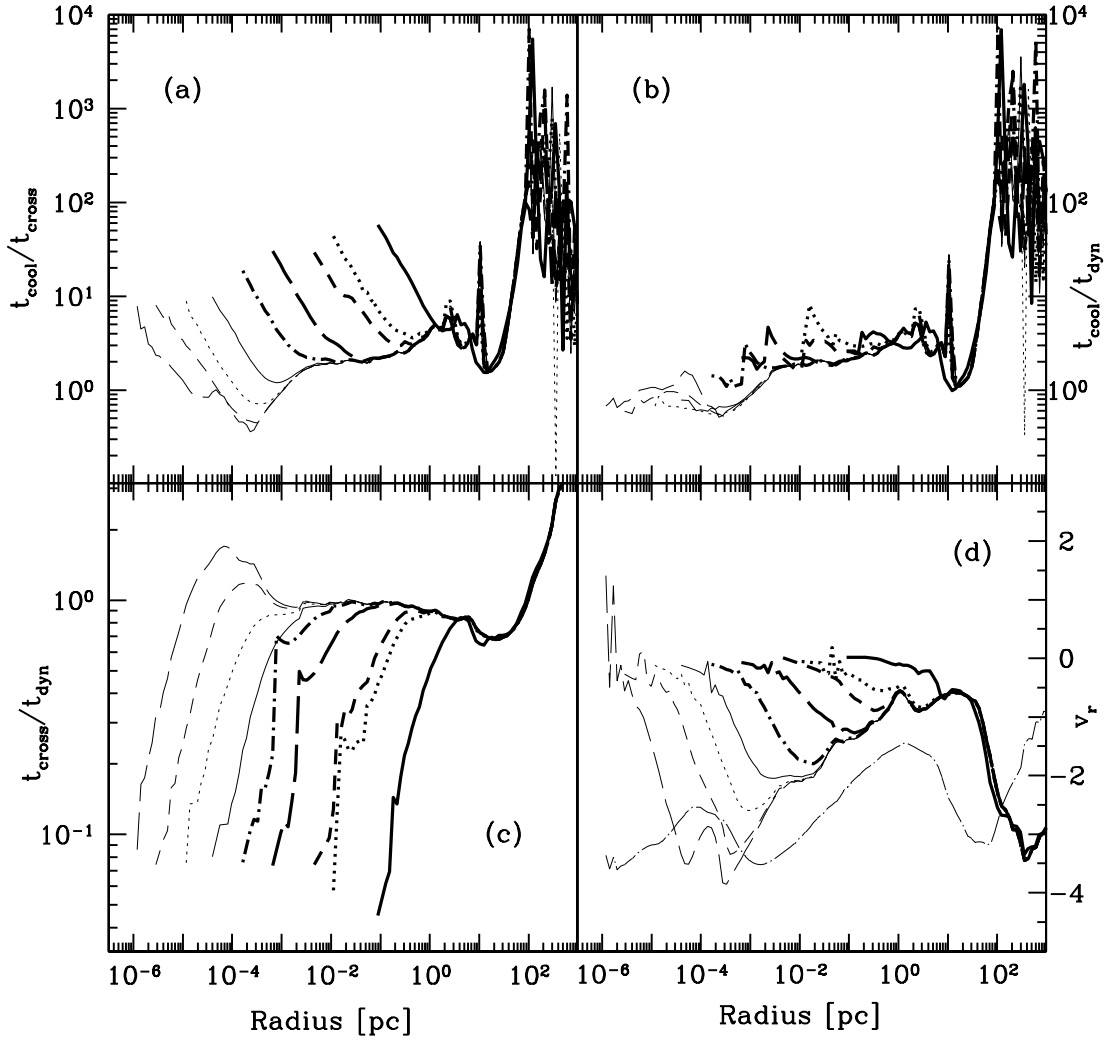}
\end{center}
\caption{
Evolution of the ratio of gas cooling time to sound crossing time (a),
gas cooling time to system dynamical time (b), sound crossing time
to system dynamical time (c) and baryon radial velocity 
(d) as a 
function of radius of a representative primordial protostar.  These quantities are
mass-weighted and spherically-averaged, and the lines correspond to the same times
as in Figure~\ref{fig.rep.panel1} and are of the same
simulation.
The black dot-dashed line in the plot of radial velocity as
a function of radius is the sound speed calculated using the local baryon temperature in
each radial bin at the
last simulation timestep. }
\label{fig.rep.panel3}
\end{figure}

\clearpage
Figures~\ref{fig.rep.panel1} through~\ref{fig.rep.panel3} show the evolution of
radial profiles of several spherically averaged, mass-weighted baryon quantities of a 
representative primordial protostar from approximately the onset of halo collapse
until the formation of a fully molecular core.  The halo begins its collapse
at $z=18.05$ (approximately $2.04 \times 10^8$ years after the Big Bang) and
ends its collapse $6.294 \times 10^6$ years later, at $z=17.67$.  
Figure~\ref{fig.rep.panel1} shows the spherically-averaged baryon number density,
temperature, and enclosed mass as a function of radius, and the specific angular momentum
of the baryon gas as a function of enclosed mass.  Figure~\ref{fig.rep.panel2}
shows the molecular hydrogen fraction, electron fraction, $H^-$ fraction and circular
velocity as a 
function of radius. 
Figure~\ref{fig.rep.panel3} shows the ratios of gas cooling time to sound
crossing time, cooling time to system dynamical time, sound crossing
time to dynamical time, and radial velocity as a function of radius.  The 
lines in all of these plots
are chosen such that the same line thickness and type corresponds to the same time
in each panel.

We begin to follow the evolution of the halo at $z = 18.05$, 
when the central hydrogen number density has reached $n_H \sim 10^3$ particles per cubic centimeter
(black solid line in all plots).  
This corresponds to a core with a radius of $\sim 1$ parsec and a mass of
a few thousand solar masses, which is accreting gas at a significant rate.
The molecular hydrogen fraction within this
core is slightly less than $10^{-3}$ but is still enough to rapidly cool the center of
the halo to $\sim 200$ Kelvin at a cooling rate proportional to the square
of the gas density.  The gas cannot cool below this temperature because of
the sharp decrease in the cooling rate of molecular hydrogen below $\simeq 200$ Kelvin.
This core is the high-redshift equivalent of a molecular cloud core.  The halo ``loiters''
for approximately six million years as the amount of molecular hydrogen is slowly built up
to a mass fraction of a few times $10^{-3}$ and the central density increases.  As can be seen from
Figure~\ref{fig.rep.panel3}, the cooling time in the halo core is 
significantly larger than both the dynamical
time and sound crossing time at this epoch, so the core can be quite reasonably said to be 
quasistatically evolving.  It is useful to compare the late-time evolution of a Population
III minihalo to the \markcite{1977ApJ...214..488S}{Shu} (1977) isothermal sphere model of cloud collapse,
which is most commonly applied to quasistatic cores in galactic molecular clouds.

As the
gas density passes roughly $n_H \sim 10^4$ cm$^{-3}$ the ro-vibrational levels of H$_2$ are populated
at their thermodynamic equilibrium values and the cooling rate 
becomes independent of density, which corresponds
to an increase in gas temperature with increasing density (as can be seen by the blue and green
solid lines in the temperature vs. radius plot in Figure~\ref{fig.rep.panel1}).  As the temperature
increases the cooling rate again begins to rise, leading to an increase in the inflow velocities of 
gas.  This is easily explained by the Shu model, as the infall of gas onto the quasistatic halo is
moderated by the sound speed, and the accretion rate $\dot{m}$ scales as the sound speed cubed.
Examination of the plot of enclosed mass vs. radius in Figure~\ref{fig.rep.panel1}
shows that at this point the enclosed gas mass has exceeded the Bonnor-Ebert critical mass, 
which is defined as $M_{BE} = 1.18 M_\odot (c_s^4/G^{3/2}) P_{ext}^{-1/2}$, where $c_s$ is
the local sound speed and $G$ is Newton's constant.  This is the critical mass
at which an isothermal sphere of gas with an external pressure $P_{ext}$ becomes gravitationally
unstable and undergoes collapse.  This occurs in this halo at a mass scale of $\sim 1000$~M$_\odot$.

When the central density of the cloud core becomes sufficiently large ($n_H \sim 10^8$ cm$^{-3}$)
the three-body H$_2$ formation process takes over, resulting in a rapid increase in the molecular
hydrogen fraction from a few times $10^{-3}$ to essentially unity.  The halo densities
also cause a rapid drop in the $H^-$ and electron fractions, though the dominance of the 
3-body channel makes the $H^-$ channel for molecular hydrogen formation unimportant.  This causes 
a huge
increase in the cooling rate, which results in a rapid drop in temperature of the center of
the halo, allowing it to contract and causing an increase in central density of 
$n_H \sim 10^{15}$~cm$^{-3}$ in only 
$\sim 2 \times 10^4$ years.  The infall velocity of gas to the center of the halo
peaks at the same radius as the temperature, and drops within that radius (r~$\simeq 10^{-3}$~pc).  
At a mass
scale of $\sim 1$~M$_\odot$ a protostellar cloud core forms which is completely molecular and
has gas accreting onto it supersonically.  At this point the optical depth of the
halo core becomes close to unity to molecular hydrogen ro-vibrational line emission, 
so we terminate the simulation because 
the assumption of optically thin radiative cooling used in our code is no longer correct.  Note
also the change from supersonic back to subsonic accretion at very small radii 
in Figure~\ref{fig.rep.panel3} (d) is 
not an accretion shock onto a protostar -- rather, it is a numerical effect due to the flow 
converging to a single point in a grid cell with finite resolution.

Panel (d) in Figures~\ref{fig.rep.panel1} through~\ref{fig.rep.panel3} 
shows the dynamical evolution of the collapsing halo.  These plots
show the spherically-averaged, mass-weighted, specific angular momentum as 
a function of enclosed baryon mass and the spherically-averaged, mass-weighted, 
baryon radial velocity
and circular velocity as a function of radius.  Though the angular momentum of the gas within
the halo core (m$_{b} \la 10^3$~M$_\odot$) is conserved overall, there is clear evidence
for transport of angular momentum outward.  This effect can be shown to be
physical (as opposed to a numerical effect), and is due to a combination of turbulent transport
of high angular momentum gas outward and a ``segregation effect,'' where low angular momentum
gas tends to move inward during halo collapse. A detailed discussion of this issue is deferred
to the second paper in this series (O'Shea, Norman \& Li 2006, in preparation).

The thin dot-dashed lines in the plots of spherically-averaged baryon infall velocity and
cylindrically-averaged (in shells along the axis of rotation) baryon circular velocity (the
bottom right panel in Figure~\ref{fig.rep.panel2}
correspond to the locally-determined sound speed and Keplerian orbital velocities at the final timestep,
respectively.  The infall of gas is locally subsonic until very late times, when the core has reached
densities of $n_H \sim 10^{12}$~cm$^{-3}$ and an H$_2$ fraction approaching unity.  The infall velocities then
become locally supersonic, but decrease rapidly within this radius because 
a protostellar cloud core has formed which the gas is accreting onto, and has
formed a standing shock.  At no point in the simulated evolution is the gas within the halo core 
rotationally-supported: the initial distribution of gas velocities is such that gas in the core has 
very little angular momentum, and what little exists is efficiently transported outward.

It is useful to examine the relevant chemical, thermal and dynamical timescales of the 
collapsing halo.  The ratios of cooling time to sound crossing time 
(calculated as a mass-weighted mean in spherically averaged radial shells) 
as a function of radius, cooling time
to dynamical time (calculated assuming spherical symmetry), 
and sound crossing time to dynamical time are plotted in 
Figure~\ref{fig.rep.panel3}.  Within the core of the halo ($r \sim 1$ parsec) the sound
crossing time ($t_{cross}$) is slightly less than the dynamical time ($t_{dyn}$) 
for the majority of the evolution time
of the halo, while the cooling time ($t_{cool}$) is somewhat longer than both of these timescales, 
but generally only by a factor of a few.  If $t_{cross} \ll t_{dyn}$ 
the halo is stable against collapse because the halo can easily equilibrate its pressure to compensate 
for collapsing gas.  If $t_{cross} \gg t_{dyn}$, the system cannot come into equilibrium and is
in free-fall.  In this case, $t_{cross} \approx t_{dyn} < t_{cool}$, and the system is undergoing 
a quasistatic collapse.  This can also be seen by examining the evolution of the 
radial infall velocity as a function of radius in Figure~\ref{fig.rep.panel3}, where the radial
infall velocity is subsonic until the very last portion of the core's evolution, when it becomes
locally supersonic.  This is triggered by a dramatic increase in the molecular hydrogen fraction, 
and a corresponding rapid decrease in the cooling time.  In the center of the halo at the last few
data outputs, the cooling time becomes shorter than both the dynamical time and sound crossing time, 
creating a situation where gas is essentially free-falling onto the central protostar.

As in ABN02, we carefully examine the collapsing protostellar cloud core for signs of 
fragmentation.  This might be
expected due to chemothermal instabilities caused by the rapid formation of molecular hydrogen via
the 3-body process and the resulting rapid increase in cooling rate~\markcite{1983MNRAS.205..705S}({Silk} 1983).
  However, the sound crossing
time within the core is less than the H$_2$ formation timescale until the last output time, 
allowing mixing to take place and preventing the formation of large density contrasts.  By the 
time that the H$_2$ formation timescale becomes shorter than the sound crossing time, the core
is fully molecular and therefore stable against this chemothermal instability.
For an alternate explanation, see~\markcite{2003ApJ...599..746O}{Omukai} \& {Yoshii} (2003).

\clearpage

\begin{figure}
\begin{center}
\includegraphics[width=0.7\textwidth]{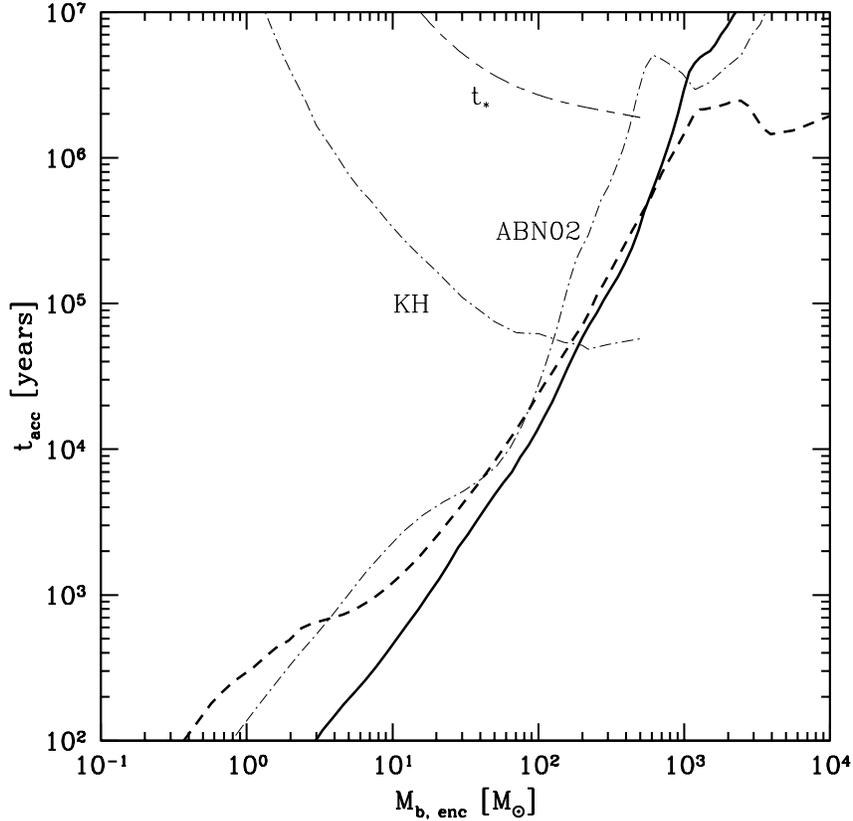}
\end{center}
\caption{
Baryon gas accretion time as a function of enclosed baryon mass 
for simulation L0\_30A.  This is defined 
as $M_{enc}/\dot{m}$, where $M_{enc}$ is the enclosed baryon mass
and $\dot{m} \equiv 4 \pi r^2 \rho(r) v(r)$, with $\rho(r)$ and 
$v(r)$ being the baryon density and velocity as a function of radius,
and $v(r)$ defined as being positive towards the center of the halo.
The thick solid line is the baryon accretion time for this simulation.  The 
thick dashed line is the accretion time as determined by taking the accretion
rate from the Shu isothermal collapse model, $\dot{m}_{Shu} = m_0 c_s^3 / G$, where
$m_0$ is a dimensionless constant of order unity, $c_s$ is the sound speed, and
$G$ is the gravitational constant.  This value of $\dot{m}$ is calculated in each
bin and the accretion time is plotted as $M_{enc} / \dot{m}_{Shu}$.
The thin line with the label ``KH'' is the Kelvin-Helmholtz time for a Population
III star with a mass identical to the enclosed mass, as calculated from 
the results given by Schaerer.  The thin line labeled ``$t_*$'' is the 
lifetime of a massive primordial star assuming no significant mass loss,
as calculated by Schaerer.  Note that this fit only covers the mass range
$5-500$~M$_\odot$, but $t_* \simeq$ constant for $M_* > 500$~M$_\odot$.
The thin dot-short dashed line labeled ``ABN02'' is the baryon accretion time for the result 
in ABN02.  
The plot here corresponds to the last output dataset,
corresponding to the thin, long-dashed line in Figures~\ref{fig.rep.panel1} 
through~\ref{fig.rep.panel3}.}
\label{fig.rep.accrete}
\end{figure}

\clearpage
As discussed previously, at the time when the simulation is stopped due to a breakdown in 
the assumption of optically thin radiative cooling at the center of the protostellar cloud ,
a fully-molecular core with a mass of $\sim 1$~M$_\odot$ has formed and is accreting
gas supersonically.  The spherically-averaged accretion time at the last output timestep, 
plotted
as a function of enclosed gas
mass, is shown as the thick solid line in Figure~\ref{fig.rep.accrete}.  The accretion time
is defined as $t_{acc} \equiv M_{enc}/\dot{m}$, where $M_{enc}$ is the enclosed baryon mass
and $\dot{m} \equiv 4 \pi r^2 \rho(r) v(r)$, with $\rho(r)$ and 
$v(r)$ being the spherically-averaged, mass-weighted baryon density and velocity as a function of radius,
and $v(r)$ defined as being positive towards the center of the halo.  The thick dashed line is
the accretion time as determined by taking the local accretion rate from the Shu isothermal
collapse model, $\dot{m}_{Shu} = m_0 c_s^3 / G$, where
$m_0$ is a dimensionless constant of order unity (we use $m_0 = 0.975$, as derived by Shu), 
$c_s$ is the sound speed, and
$G$ is the gravitational constant.  This value of $\dot{m}$ is calculated in each
bin and the accretion time is plotted as $M_{enc} / \dot{m}_{Shu}$.
The dot-long dashed line is the Kelvin-Helmholtz time for a Population
III star with a mass identical to the enclosed mass calculated assuming a star with
a main sequence radius and luminosity as given by \markcite{2002A&A...382...28S}{Schaerer} (2002).  The
short dash-long dashed line is the lifetime of a massive primordial star as calculated
by Schaerer, assuming no mass loss.  Note that this fit only covers the mass range
$5-500$~M$_\odot$, but the lifetime of massive stars with a mass greater than $500$~M$_\odot$ is
roughly constant. 
The dot-short dashed line is the baryon accretion time for the result 
in ABN02.

We can see that the accretion rates in our model and in ABN02 are well-approximated
by the Shu model for $M_{enc} \lesssim 10^3$~M$_\odot$, which is the Bonner-Ebert 
mass scale of the collapsing core.  All three curves predict that
$100-200$~M$_\odot$ of gas will accrete within the Kelvin-Helmholtz time of the 
Population III star, implying that it will be at least this massive.  As we will see
in the next section, the similarity of these curves is somewhat fortuitous, as we find
substantial variation of accretion times within our suite of models.
\clearpage

\clearpage

\section{Comparison of multiple realizations}\label{compare-real}

Another important issue relating to studies of the formation of Population
III stars in a cosmological context is the consistency of the results over
a range of simulation parameters.  As
discussed in the introduction to this article, previously published
high dynamical range calculations of Pop III star formation have concentrated
upon a single cosmological realization.  While this is an important first step, it
neglects possible systematic effects and also allows for 
error due to small number statistics.

In this section we attempt to address some of these issues.  Twelve simulations
are set up as described in Section~\ref{simsetup}.  Each simulation has the same cosmological 
parameters but a 
different cosmological realization (i.e. large scale structure).  Four
simulations in each of three box sizes ($0.3, 0.45$, and~$0.6$~h$^{-1}$~comoving Mpc) 
are performed, with the results shown in Figures~\ref{fig.comp.panel1} 
through~\ref{fig.accrete.panel3}.

\subsection{Halo bulk properties}\label{sec:halo-bulk}

\begin{figure}
\begin{center}
\includegraphics[width=0.9\textwidth]{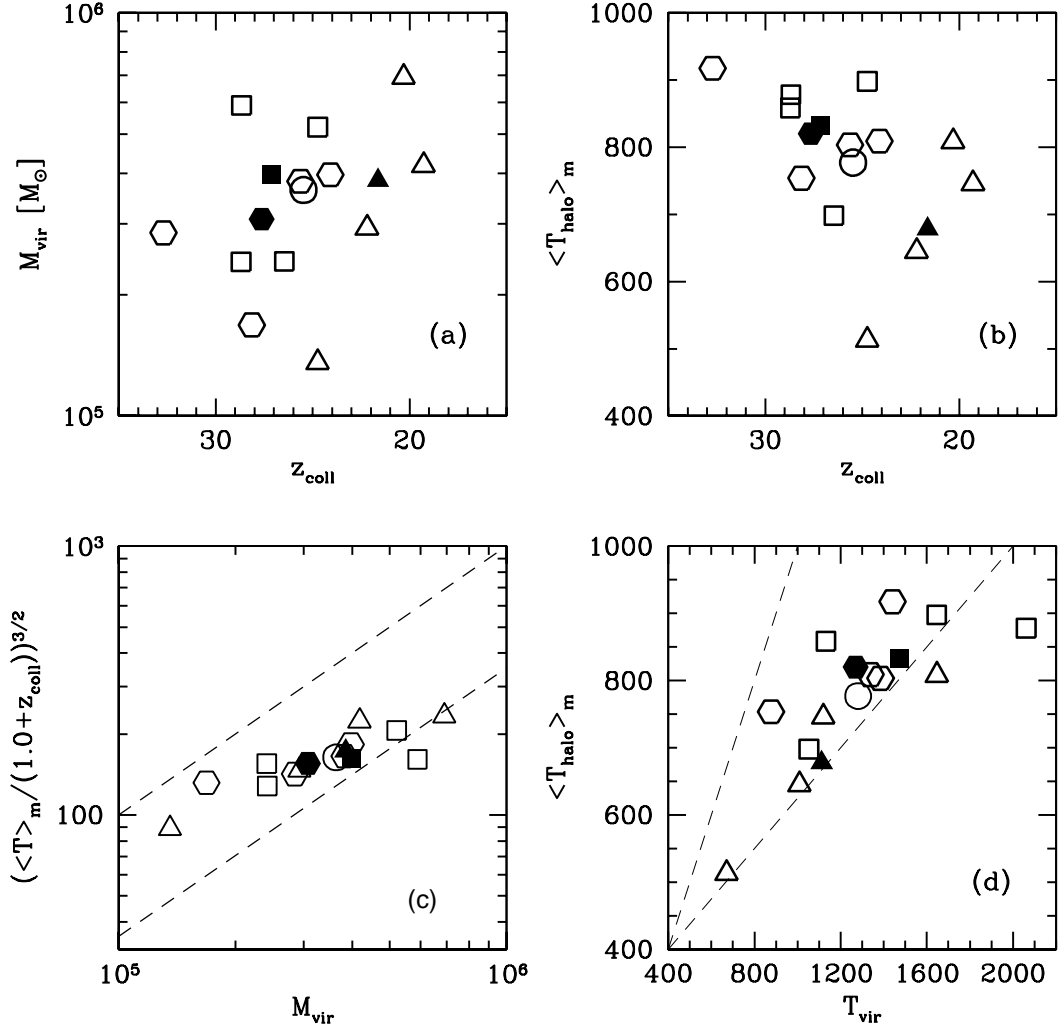}
\end{center}
\caption{
Plots of basic halo quantities for 12 different cosmological random realizations.
Panel (a):  halo virial mass vs. protostar collapse redshift.  Panel (b):
mean mass-weighted halo baryon temperature vs. collapse redshift.  Panel (c):
($(<T>_m/(1+z))^{3/2}$ vs. halo virial mass.  Panel (d):  halo mean baryon
temperature vs. halo virial temperature.  In each plot, open triangles, squares and hexagons
correspond to simulations with 0.3 h$^{-1}$~Mpc, 0.45 h$^{-1}$~Mpc and 0.6 h$^{-1}$~Mpc 
comoving box size, respectively.  The filled shapes corresponds to the 
average value for simulations with that box size.  The open circle corresponds
to the average for all simulations together.  The upper and lower dashed lines
in panels (c) and (d) correspond to the scaling relation one would expect if  $<T>_m$ = $T_{vir}$
and $<T>_m$ = 0.5 $T_{vir}$, respectively.  
}
\label{fig.comp.panel1}
\end{figure}

\begin{figure}
\begin{center}
\includegraphics[width=0.9\textwidth]{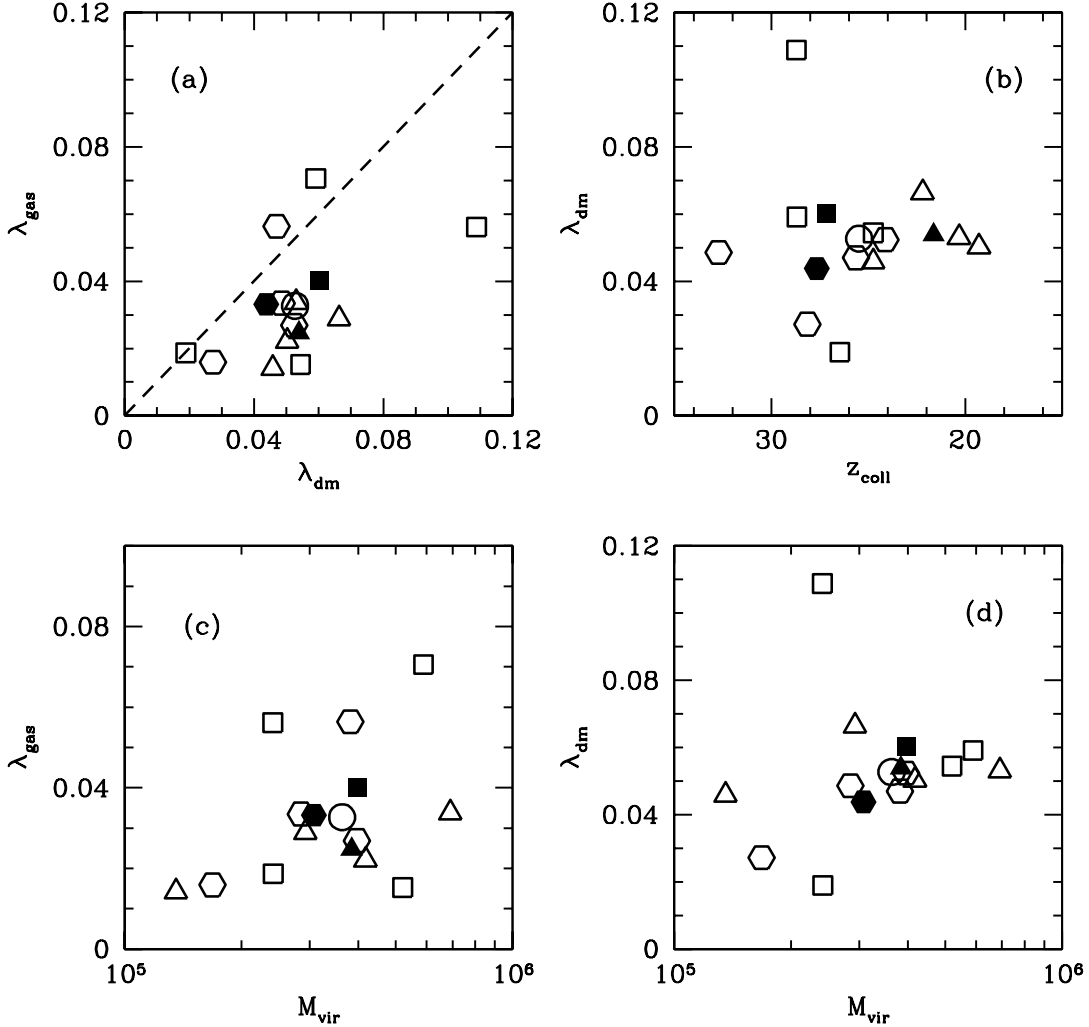}
\end{center}
\caption{
Plots of basic halo quantities for 12 different cosmological random realizations.
Panel (a):  gas spin parameter vs. dark matter spin parameter.  Panel (b): dark 
matter spin parameter vs. halo collapse redshift.  Panel (c):  gas
spin parameter vs. halo virial mass.  Panel (d):  dark matter spin parameter vs.
halo virial mass.
In each plot, open triangles, squares and hexagons 
correspond to simulations with 0.3 h$^{-1}$~Mpc, 0.45 h$^{-1}$~Mpc and 0.6 h$^{-1}$~Mpc 
comoving box sizes, respectively.  The filled shapes corresponds to the 
average value for simulations with that box size.  The open circle 
corresponds to the average for all simulations together.
}
\label{fig.comp.panel2}
\end{figure}

\begin{figure}
\begin{center}
\includegraphics[width=0.9\textwidth]{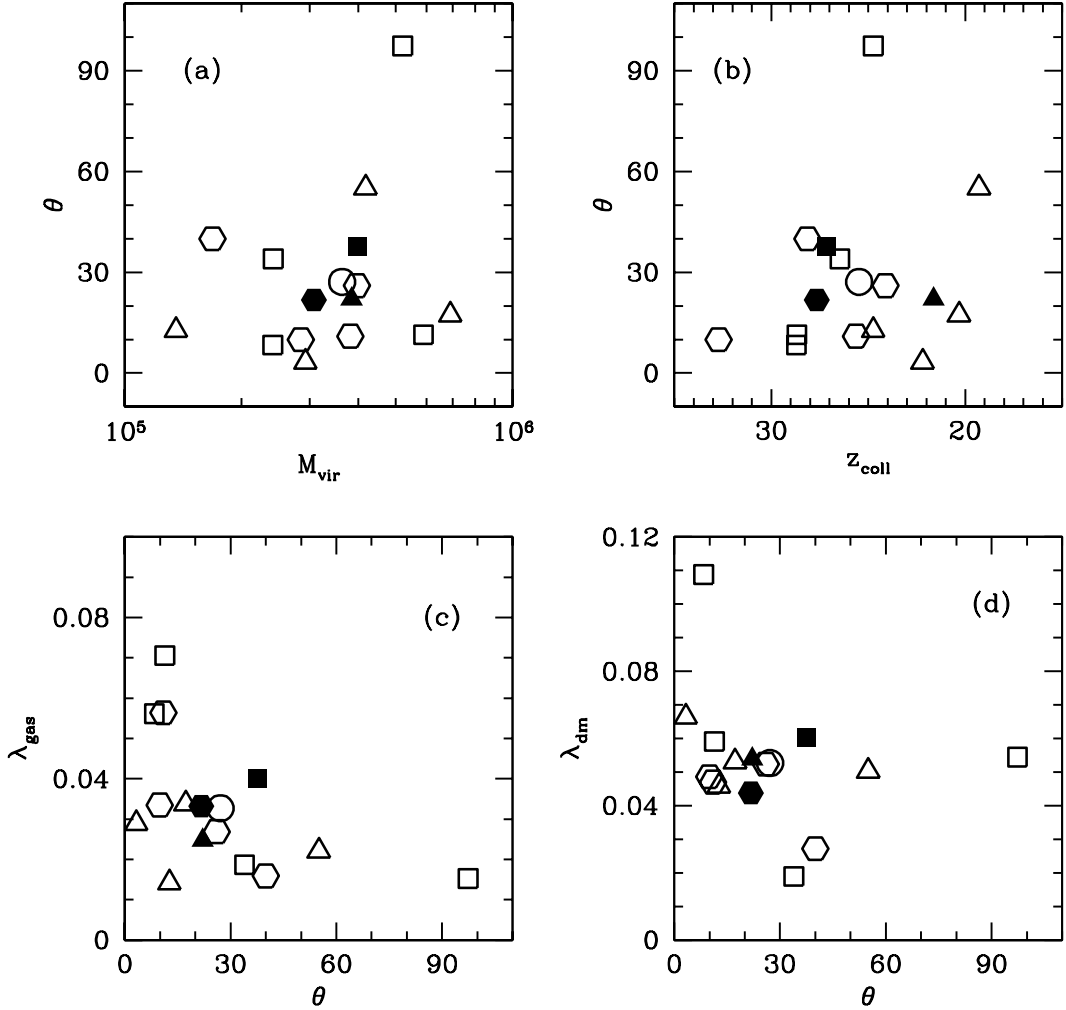}
\end{center}
\caption{
Plots of basic halo quantities for 12 different cosmological random realizations.
Panel (a):  Theta (angle between gas and dark matter angular momentum vectors)
vs. halo virial mass.  Panel (b):  theta vs. halo collapse redshift.
Panel (c):  gas spin parameter vs. theta.  Panel (d):  dark matter
spin parameter vs. theta.
In each plot, open triangles, squares and hexagons 
correspond to simulations with 0.3 h$^{-1}$~Mpc, 0.45 h$^{-1}$~Mpc and 0.6 h$^{-1}$~Mpc 
comoving box sizes, respectively.  The filled shapes corresponds to the 
average value for simulations with that box size.  The open circle 
corresponds to the average for all simulations together.}
\label{fig.comp.panel3}
\end{figure}

Figures~\ref{fig.comp.panel1} - \ref{fig.comp.panel3} display several
mean properties of the halos.  In each panel information
for each simulation is plotted as an open shape which assigned according
to box size as described in the figure captions.  The filled-in shapes correspond to mean 
values for all simulations of a given
box size (with shapes again corresponding to the box size), and the open circle corresponds
to the mean of all twelve of the simulations together.  All quantities discussed
in this section can be found in Tables~\ref{table.meaninfo-1} and~\ref{table.meaninfo-2}.

Panel (a) of Figure~\ref{fig.comp.panel1} shows the virial mass of
each halo calculated using the virial density fitting function of~\markcite{1998ApJ...495...80B}{Bryan} \& {Norman} (1998)
at the time of protostellar cloud collapse plotted against the redshift of collapse.  
Though there is a large amount of scatter in virial mass overall,
with the smallest halo having a mass of $1.36 \times 10^5$~M$_\odot$ and the 
largest $6.92 \times 10^5$~M$_\odot$, the average 
virial mass in each box size is virtually identical.  The mean virial mass of
all twelve of the halos is $3.63 \times 10^5$~M$_\odot$, which is significantly 
lower than the halo mass of $7 \times 10^5$~M$_\odot$ in ABN02 and the minimum mass of
$5 \times 10^5$~h$^{-1}$~M$_\odot$ found by~\markcite{2003ApJ...592..645Y}{Yoshida} {et~al.} (2003) 
(note that the value found
by ABN02 is from a different cosmology with much higher values of $\Omega_m$ and $\Omega_b$,
yet is still within the scatter of halo virial masses found in this work).  In contrast to the
virial mass, there is an apparent trend in earlier collapse time (large collapse redshift)
as a function of box size, with the 0.45 and 0.6 h$^{-1}$ Mpc boxes collapsing at a
mean redshift of $z \simeq 27.5$ and the 0.3 h$^{-1}$ Mpc boxes collapsing at a mean
redshift of $z \simeq 22$.  This can be understood as a result of the larger boxes
sampling a much larger volume of the universe.  The first halo to collapse in the larger
volume tends to be a statistically
rarer density fluctuation that collapses at a higher redshift.  A consequence
of this can be seen in panel (b) of Figure~\ref{fig.comp.panel1}, which 
shows the mean baryon temperature in each halo as a function of collapse redshift.
As with the plot of virial mass vs. collapse redshift, there is a significant amount
of scatter in the results, but a clear trend of increased overall halo temperature 
with collapse redshift is apparent.  This is explainable in terms of virial properties
of the halo.  The virial temperature scales as $T_{vir} \sim M_{vir}^{2/3} (1+z)$, so halos 
with a constant virial mass will be hotter at higher redshifts, as observed in this plot.
\clearpage
\clearpage
The virial scaling of the mean halo baryon temperature vs. halo virial mass ($(<T>_m/(1+z))^{3/2}$ vs.
$M_{vir}$) is plotted in panel (c) of Figure~\ref{fig.comp.panel1}, and the mean 
halo temperature versus
the halo virial temperature is plotted in panel (d).  The upper and lower 
dashed lines in both panels  correspond to the scaling relation one would expect if  $<T>_m$ = $T_{vir}$
and $<T>_m$ = 0.5 $T_{vir}$, respectively.  Since the actual
observed mean halo temperatures are significantly lower than the virial temperature, as shown in panel (d), 
one can infer that 
radiative cooling plays a significant role in the condensation of gas
in these halos despite the generally poor cooling properties of 
primordial gas at low temperature.
If radiative cooling were completely unimportant the mean halo baryon temperature would be 
approximately the virial temperature for all halos.  Examination of panels (c) and (d) together show that
the temperature evolution of the halos is consistent with virial scaling, given the importance of radiative 
cooling.

Figure~\ref{fig.comp.panel2} shows the relationship of the angular momentum in 
the halos with various quantities.  The angular momentum of a cosmological halo
can be described as a function of the dimensionless spin parameter,  
$\lambda \equiv J |E|^{1/2} / G M^{5/2}$, where J is angular momentum, E
is the total energy, G is the gravitational constant and M is the halo mass.  This is roughly
equivalent to the ratio of the angular momentum of the halo to the angular momentum needed
for the halo to be rotationally supported. 
\clearpage
\clearpage
Panel (a) of Figure~\ref{fig.comp.panel2} shows the gas and dark matter spin
parameters plotted against each other for the halo in each simulation that
forms the primordial protostellar core, at the epoch of collapse.  The mean value
of the dark matter spin parameter is approximately 0.05.  The gas spin parameter tends to have
somewhat lower values, with a mean of $\simeq 0.035$, but, 
considering the number of halos examined, is consistent 
with the dark matter spin parameter.  There appears to be some overall positive correlation
between the dark matter and baryon spin parameters (e.g. halos with higher overall dark matter
angular momentum tend to have higher overall baryon angular momentum) but there is considerable
scatter.  In all cases the spin parameters are much less than one, which suggests that the 
halos have little overall angular momentum.  The measured range and distribution of the spin 
parameters agree well with previous analytical and numerical results which predict typical 
values for the spin parameter of $0.02-0.1$, with a mean value of $\simeq 0.05$ 
~\markcite{1987ApJ...319..575B,2001ApJ...557..616G, 2002ApJ...576...21V, 2003ApJ...592..645Y}({Barnes} \& {Efstathiou} 1987; {Gardner} 2001; {van den Bosch} {et~al.} 2002; {Yoshida} {et~al.} 2003).

Panel (b) of Figure~\ref{fig.comp.panel2} plots dark matter spin parameter
vs. collapse redshift of the halo.  There is no evidence that the collapse redshift systematically 
depends
on the spin parameter.  Panels (c) and (d) of Figure~\ref{fig.comp.panel2} plot the 
baryon and dark matter spin parameters against the halo virial mass.  As with the other quantities
examined, there is considerable scatter in the distributions, but no evidence for a clear relationship
between halo virial mass and either gas or dark matter spin parameter.  In all of the panels of this
figure there is no evidence for any systematic effect due to box size or collapse redshift.

Figure~\ref{fig.comp.panel3} plots the angle between the overall dark matter and baryon
angular momentum vectors ($\theta$) versus several different quantities.  Panel (a)
plots $\theta$ vs. halo virial mass at the time of formation of the Pop III protostar in each
halo.  Overall, the average value for $\theta$ is approximately 25 degrees, which is consistent with
recent numerical simulations~\markcite{2002ApJ...576...21V, 2003ApJ...592..645Y}({van den Bosch} {et~al.} 2002; {Yoshida} {et~al.} 2003).  
There is a great deal of scatter in $\theta$, which is also 
consistent.  There is no evidence for correlation between $\theta$ and halo virial mass.  Panel (b)
 plots $\theta$ vs. collapse redshift for each simulation, and panels (c) and (d)
plot the gas and dark matter spin parameters vs. $\theta$, respectively.
There appears to be no correlation between $\theta$ and collapse redshift or the
gas or dark matter spin parameters, and no evidence of there being any systematic effect due to
box size.

\clearpage
\clearpage

\subsection{Halo radial profiles}\label{sec:halo-radial}

\begin{figure}
\begin{center}
\includegraphics[width=0.9\textwidth]{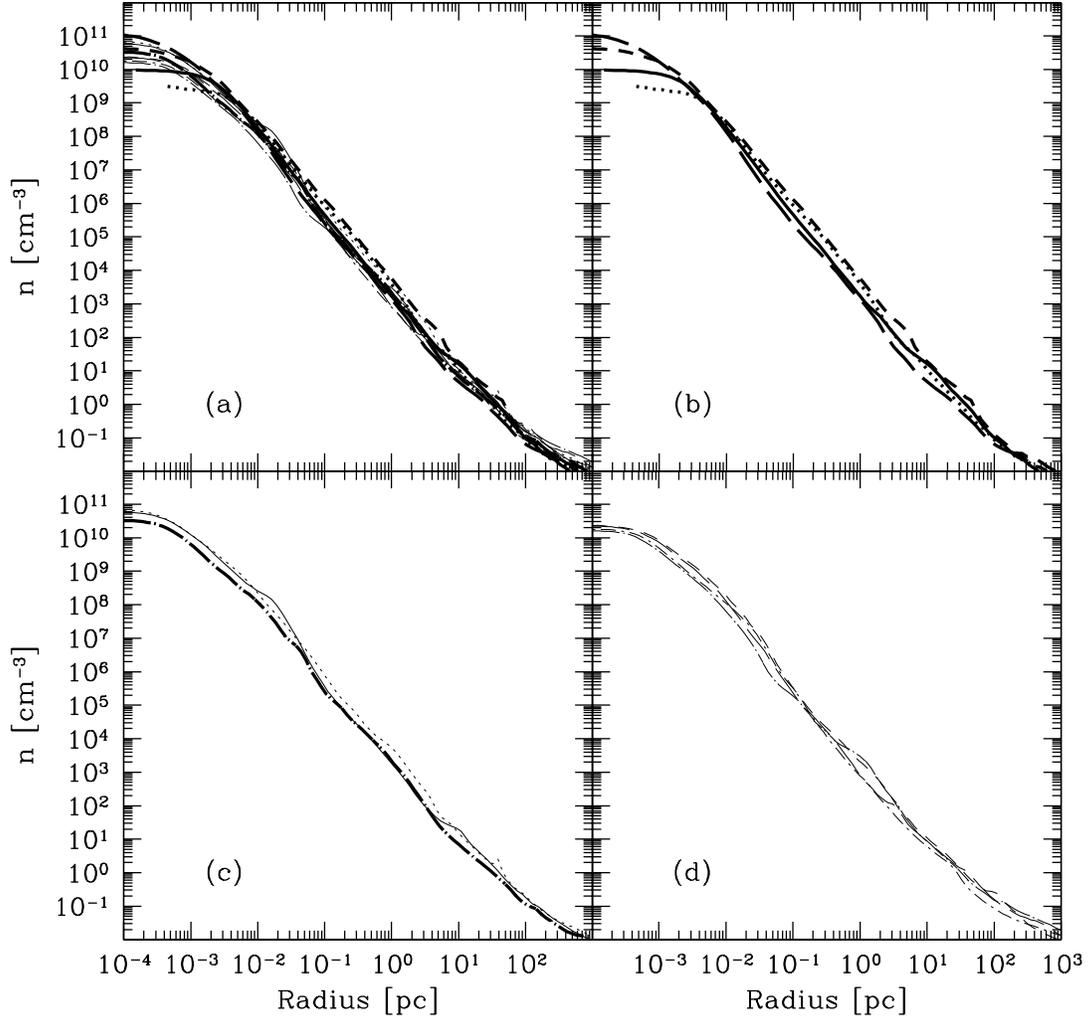}
\end{center}
\caption{
Panel (a): mass weighted, spherically-averaged baryon number density  as a function of radius 
for 11 different cosmological random realizations,
chosen at an output time where peak baryon density values are approximately the same.  (b-d): Density profiles
separated according to box size.  (b): 0.3 h$^{-1}$~Mpc, (c) 0.45 h$^{-1}$~Mpc, (d) 0.6 h$^{-1}$~Mpc (comoving).
One of the 0.45 h$^{-1}$~Mpc simulations has been omitted since the simulation crashed 
before reaching a comparable density value.  Line thicknesses and types correspond to individual
simulations.  
L0\_3A:  Thick solid line.
L0\_3B:  Thick dotted line.
L0\_3C:  Thick short-dashed line.
L0\_3D:  Thick long-dashed line.
L0\_45A: Thick dot-short dashed line.
L0\_45B: Thick dot-long dashed line.
L0\_45C: Thin solid line.
L0\_45D: Thin dotted line.
L0\_6A: Thin short-dashed line.
L0\_6B: Thin long-dashed line.
L0\_6C: Thin dot-short dashed line.
L0\_6D: Thin dot-long dashed line.
}
\label{fig.comp.panel4}
\end{figure}

\begin{figure}
\begin{center}
\includegraphics[width=0.9\textwidth]{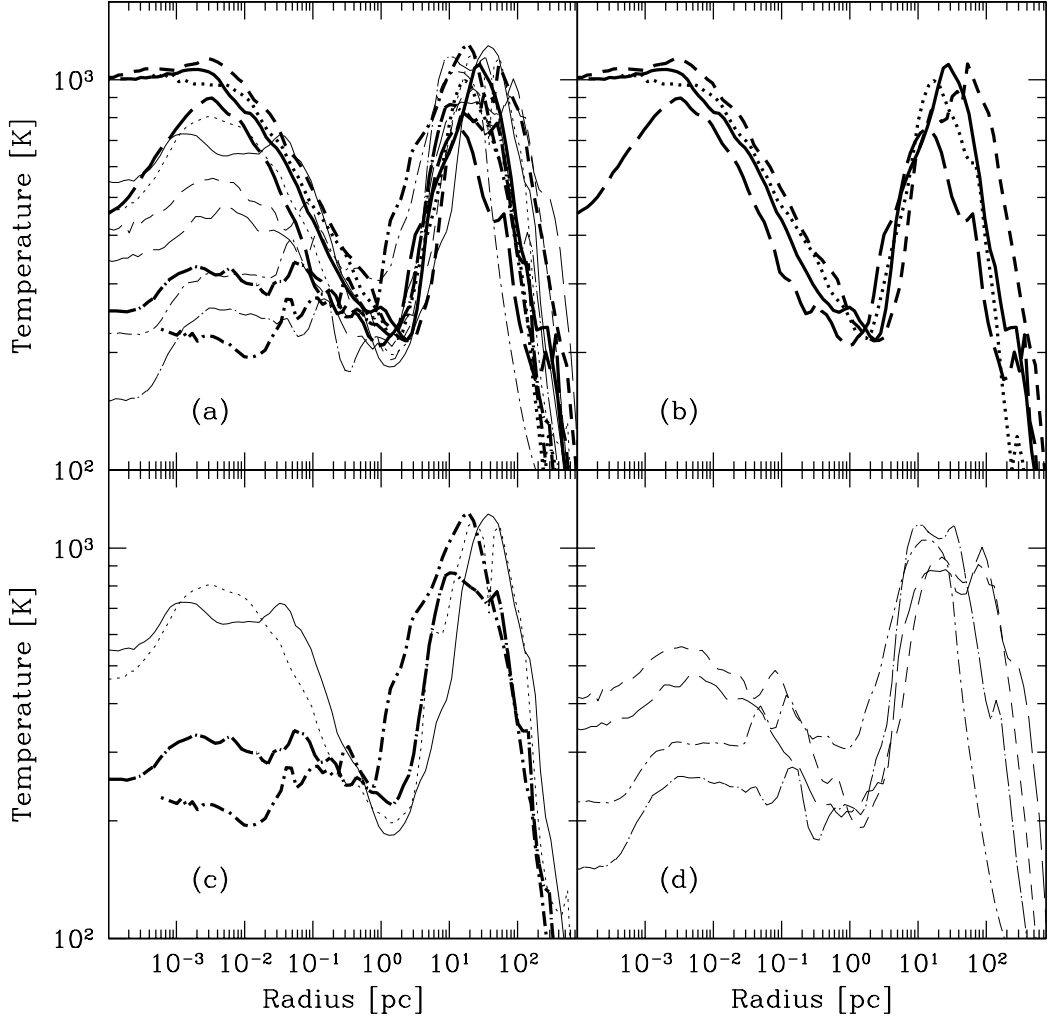}
\end{center}
\caption{
Mass weighted, spherically-averaged baryon temperature as a function of radius
for 12 different cosmological random realizations,
chosen at an output time where peak baryon density values are approximately the same.  (a) All 12 simulations, 
(b) $L_{box} =$~0.3 h$^{-1}$~Mpc, (c) $L_{box} =$~0.45 h$^{-1}$~Mpc, and (d) $L_{box} =$~0.6 h$^{-1}$~Mpc (comoving).
Line thicknesses and types are identical to those shown in Figure~\ref{fig.comp.panel4}.
}
\label{fig.comp.panel5}
\end{figure}

\begin{figure}
\begin{center}
\includegraphics[width=0.9\textwidth]{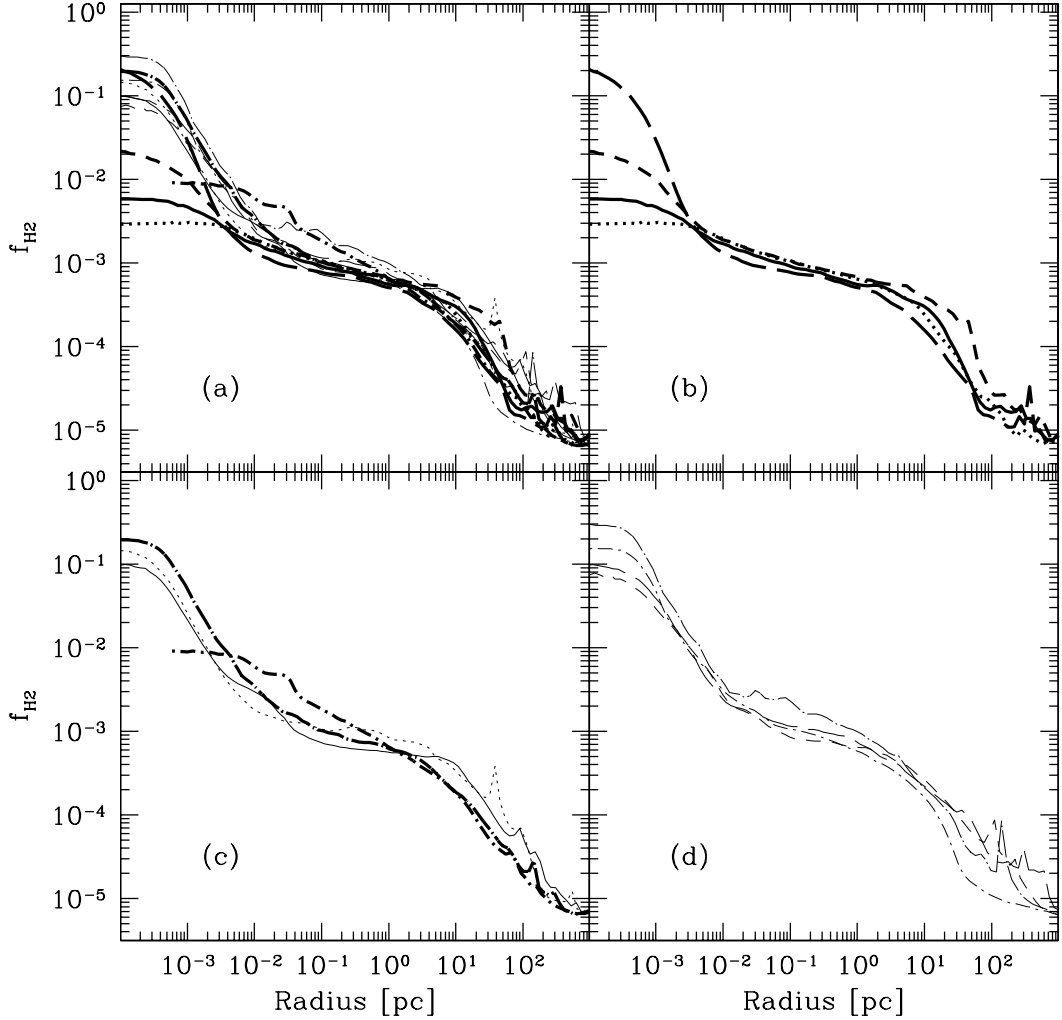}
\end{center}
\caption{
Mass weighted, spherically-averaged baryon molecular hydrogen
fraction as a function of radius
for 12 different cosmological random realizations,
chosen at an output time where peak baryon density values are approximately the same.  
(a) All 12 simulations, 
(b) $L_{box} =$~0.3 h$^{-1}$~Mpc, (c) $L_{box} =$~0.45 h$^{-1}$~Mpc, and 
(d) $L_{box} =$~0.6 h$^{-1}$~Mpc (comoving).
Line thicknesses and types are identical to those shown in Figure~\ref{fig.comp.panel4}.
}
\label{fig.comp.panel6}
\end{figure}

\begin{figure}
\begin{center}
\includegraphics[width=0.9\textwidth]{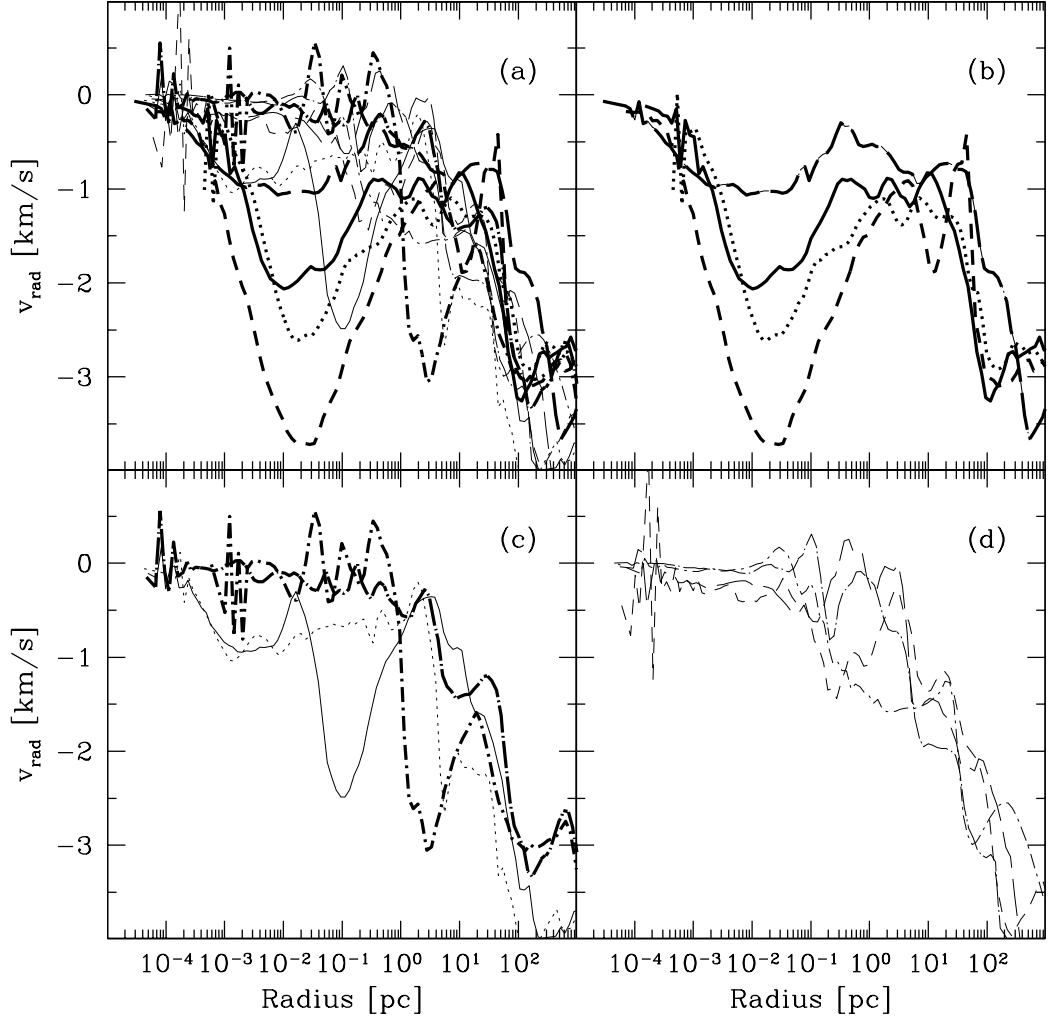}
\end{center}
\caption{
Mass-weighted, spherically-averaged baryon radial velocity as a function of 
radius for 12 different cosmological random realizations,
chosen at an output time where peak baryon density values are approximately the same.
(a) All 12 simulations, 
(b) $L_{box} =$~0.3 h$^{-1}$~Mpc, (c) $L_{box} =$~0.45 h$^{-1}$~Mpc, and 
(d) $L_{box} =$~0.6 h$^{-1}$~Mpc (comoving).
Line thicknesses and types are the same as in Figure~\ref{fig.comp.panel4}.
}
\label{fig.comp.panel7}
\end{figure}

\begin{figure}
\begin{center}
\includegraphics[width=0.9\textwidth]{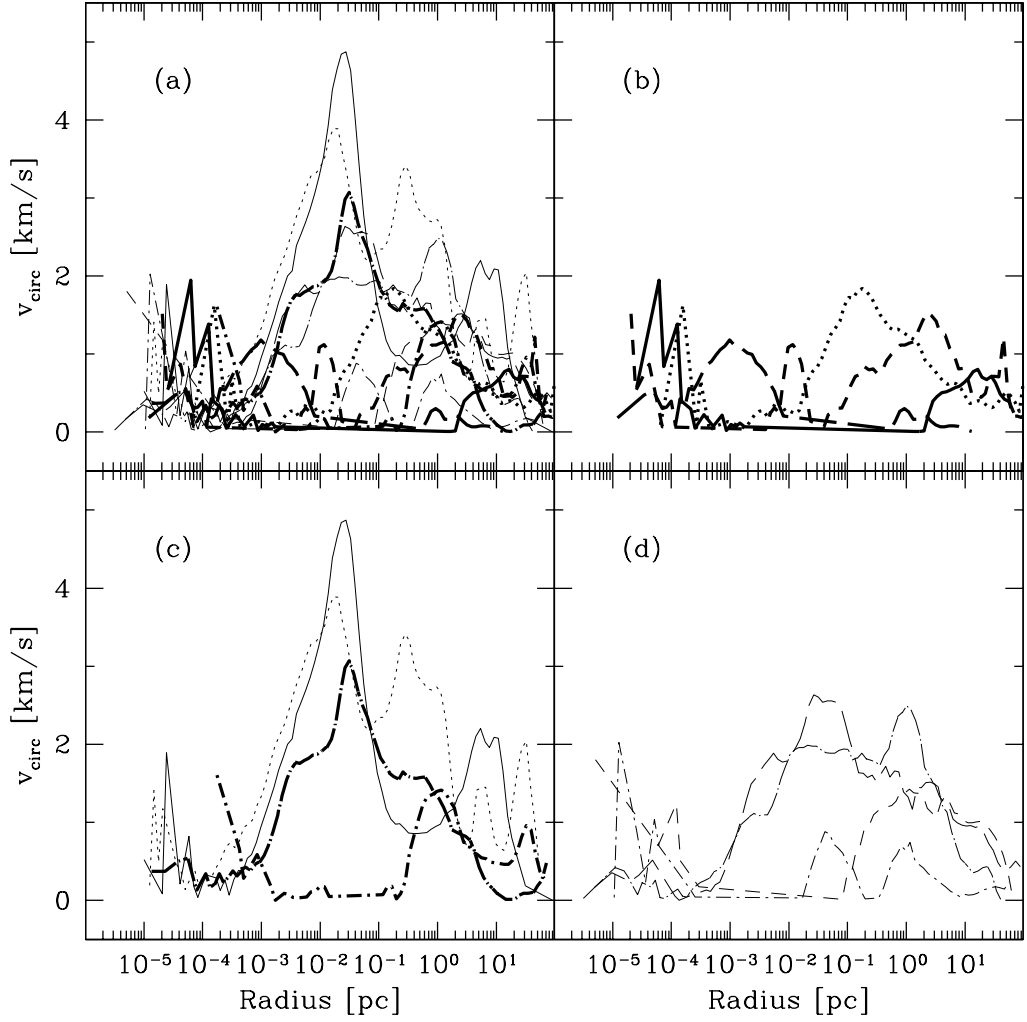}
\end{center}
\caption{
Mass weighted,
cylindrically-averaged baryon circular velocity as a function of radius 
for 12 different cosmological random realizations,
chosen at an output time where peak baryon density values are approximately the same. 
(a) All 12 simulations, 
(b) $L_{box} =$~0.3 h$^{-1}$~Mpc, (c) $L_{box} =$~0.45 h$^{-1}$~Mpc, and 
(d) $L_{box} =$~0.6 h$^{-1}$~Mpc (comoving).
Line thicknesses and types are the same as in Figure~\ref{fig.comp.panel4}.
}
\label{fig.comp.panel8}
\end{figure}

In addition to plots of mean halo properties, it is very useful to look at more detailed
information about each halo.  Figures~\ref{fig.comp.panel4} through~\ref{fig.comp.panel8}
show spherically-averaged, mass-weighted radial profiles of several baryon quantities in eleven
of the twelve simulations (one simulation crashed and could not be restarted before reaching a high
enough density).  Since the cores of the most massive halo in each simulation collapse at a 
range of redshifts, we choose to compare them at a fixed point in the halo's evolution, as 
measured by the peak central
baryon density in the protostellar cloud, which is roughly analogous to a constant point in the
evolution of the protostar.  In each of the figures discussed below, panel (a) shows
radial profiles for all of the simulations plotted together.  Panel (b) shows only
the results for the 0.3 h$^{-1}$~Mpc box, panel (c) shows only results for the
0.45 h$^{-1}$~Mpc box, and panel (d) shows only results for the 0.6 h$^{-1}$~Mpc
box.  Line of a given line type and weight
correspond to the same simulation in all figures.

Figure ~\ref{fig.comp.panel4} shows the plots of number density as a function of radius
for eleven simulations, shown at approximately the same point in their evolution.
There is remarkably little scatter in the density profiles for all of the simulations (less than
1 dex at any given radius),
and the density profiles all tend towards $\rho(r) \sim r^{-2.2}$.
It was shown by~\markcite{1968ApJ...152..515B}{Bodenheimer} \& {Sweigart} (1968) that for a cloud
of gas that is nearly isothermal and slowly rotating and which has negligible
support from a magnetic field, the subsonic evolution of the gas will 
tend to produce a $1/r^2$ density distribution as long as the thermal
pressure remains approximately in balance with the gravitational field.
In particular, \markcite{chandra39}{Chandrasekhar} (1939) showed that a molecular
cloud core which forms at subsonic speeds will tend towards the density
distribution of a singular isothermal sphere,

\begin{equation} 
\rho(r) = \frac{c_s^2}{2 \pi G r^2}
\label{eqn-sis}
\end{equation}

where $c_s \equiv (kT/m)^{1/2}$ is the isothermal sound speed, T, k, and m are the 
temperature, Boltzmann's constant, and mean molecular weight of the gas, respectively,
 and and r is the radius.  Since the halos generally have low angular momentum (as seen
in Figure~\ref{fig.comp.panel2}) and magnetic fields are completely neglected in these
simulations, it is reasonable that the density should go as $\rho(r) \sim r^{-2}$ in all of the
simulations, with the deviation from $r^{-2}$ being a result of the effective equation of state
of the primordial gas~\markcite{omukai98}({Omukai} \& {Nishi} (1998)).  The overall 
normalization of the density profiles also agrees very well.
  This can be understood as a result of the cooling properties of hydrogen gas.
Each of the halos examined in this figure has the same composition, and therefore is
cooled by the same mechanism.  Only a small amount of molecular hydrogen is needed to
cool the gas relatively efficiently, suggesting that in a halo that is in a somewhat
stable equilibrium the gas temperature at low densities should be approximately 
constant for different halos, independent of the molecular 
hydrogen fraction.  At densities above $n_H \simeq 10^4$ cm$^{-3}$ the cooling rate becomes
independent of density and the halo enters a ``loitering'' phase, where a slow buildup
of mass occurs.  Once a Bonnor-Ebert mass of gas accumulates, the halo core becomes
gravitationally unstable and begins to evolve very rapidly, 
so small differences in the initial molecular hydrogen fraction become magnified
(as discussed later in this section).  The scatter in densities at a given
radius is related to the scatter in virial masses, and
there is no systematic effect with collapse redshift.
\clearpage
\clearpage
Figure~\ref{fig.comp.panel5} shows the baryon temperature as a function of radius.  At
radii outside of $\sim1$ parsec, the temperature profiles are similar between all of 
the simulations, though halos forming at higher redshifts tend to have
a higher overall temperature.  At smaller radii there is significant scatter in core temperature
of the simulations (for a fixed density), with a systematic trend towards halos forming
in larger boxes having a lower overall core temperature.  Examination of 
Figure~\ref{fig.comp.panel6} plots molecular hydrogen mass fraction as a function 
of radius, and shows that halos which form at higher redshift have systematically
larger H$_2$ mass fractions, though this effect is much more pronounced in the core of the halo 
than in the envelope.  This difference in molecular hydrogen fraction can be understood
as a result of the overall halo temperature.  The rate at which molecular hydrogen is 
produced at low densities is limited by the availability of free electrons, as described 
by \markcite{1997ApJ...474....1T}{Tegmark} {et~al.} (1997). The mean fraction of free electrons available in the
primordial gas is a function of baryon temperature, with larger temperatures 
corresponding to larger electron fractions.  On the other hand, the rate at which 
molecular hydrogen forms via the $H^-$ channel declines at high temperatures.
Since the limiting reaction in the formation of molecular hydrogen via the $H^-$ channel is
the formation of free electrons, this reaction dominates, and it can be shown using a
simple one-zone calculation following the nonequilibrium primordial chemistry that 
molecular hydrogen production is maximized at $\sim 1000~$K.  Halos with higher overall
baryon temperatures will have systematically higher molecular hydrogen fractions.
Once the core of the halo begins to collapse to very high densities small differences in the 
molecular hydrogen fraction are amplified, resulting in a general trend towards halos with 
higher overall baryon temperatures having higher H$_2$ fractions in their cores, and thus
lower central temperatures.
\clearpage

\clearpage
Figure~\ref{fig.comp.panel7} plots the mean baryon radial velocity as a function
of radius.  This figure shows that there is a 
clear systematic effect present, where halos forming in simulations with larger
boxes having a significantly lower overall radial velocity at small radii.  This 
translates directly to a lower overall accretion rate onto protostellar forming
in halos in larger simulation volumes, which is discussed more thoroughly in
Section~\ref{sec:halo-accretion}.  As discussed in Section~\ref{repstar},
this can be understood using the Shu isothermal sphere model, where subsonic collapse of gas onto the 
core of the sphere occurs at a rate controlled by the sound speed.  Since the core temperatures are lower
overall in the large simulation volumes, this translates to a lower sound speed and overall lower 
radial infall velocity.  A lower infall velocity implies a lower accretion rate, which may have
profound implications for the Population III IMF -- and are discussed elsewhere in this work.
\clearpage
\clearpage

Figure~\ref{fig.comp.panel8} 
shows the baron circular velocity in each halo as a function of 
radius. There is a great deal of scatter in this relationship, though there is no clear trend
with simulation box size.  In all cases the overall circular velocity is significantly 
less than the Keplerian orbital velocity, which agrees with our previous observation that
the halos have generally low angular momentum.
\clearpage
\clearpage

\subsection{Protostellar accretion rates}\label{sec:halo-accretion}

\begin{figure}
\begin{center}
\includegraphics[width=0.7\textwidth]{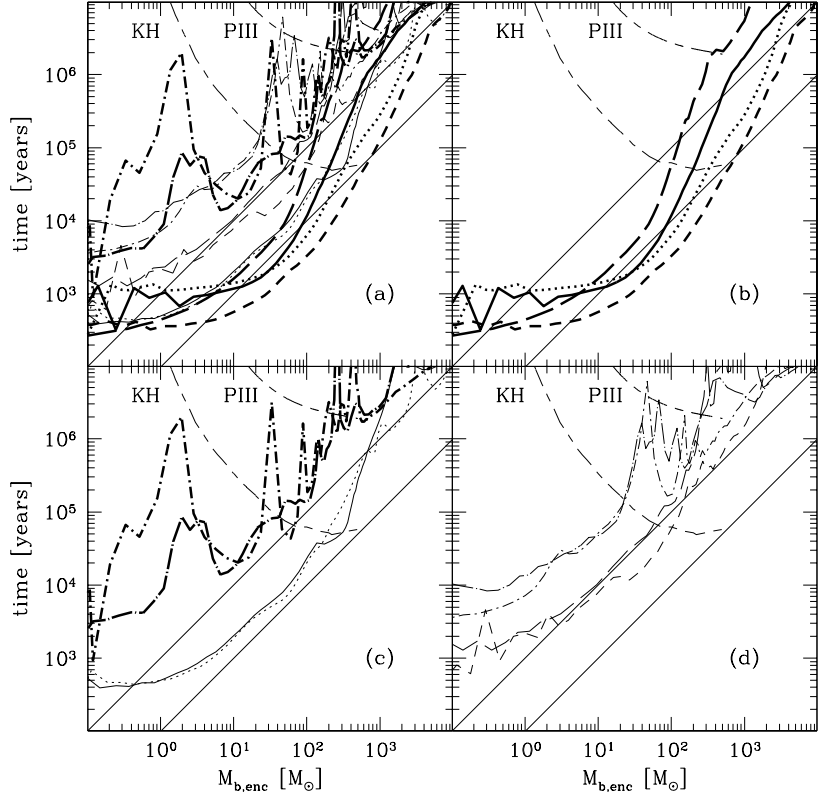}
\end{center}
\caption{
Mass-weighted, spherically averaged gas accretion time as a function of 
enclosed baryon mass for 12 different cosmological random realizations,
chosen at an output time where peak baryon density values are approximately the same.  
Panel (a):  All simulations plotted
together.  Panel (b):  0.3 h$^{-1}$~Mpc box simulations.  Panel (c):
0.45 h$^{-1}$~Mpc box simulations.  Panel (d):  0.6 h$^{-1}$~Mpc box simulations. 
Individual simulations are identified by line thicknesses and types identical to those
in Figure~\ref{fig.comp.panel4}.
The baryon accretion time is defined 
as $T_{acc} \equiv M_{enc}/\dot{m}$, where $M_{enc}$ is the enclosed baryon mass
and $\dot{m} \equiv 4 \pi r^2 \rho(r) v(r)$, with $\rho(r)$ and 
$v(r)$ being the spherically-averaged baryon density and velocity as calculated on
spherical shells,
and $v(r)$ defined as being positive towards the center of the halo.
The dot-long dashed line in each panel is the Kelvin-Helmholtz time for a Population
III star with a mass identical to the enclosed mass, as calculated from 
the results given by Schaerer.  The long dash-short dashed line is the 
lifetime of a massive primordial star assuming no significant mass loss,
as calculated by Schaerer.  Note that this fit only covers the mass range
$5-500$~M$_\odot$, by $t_* \simeq$ constant for $M_* > 500$~M$_\odot$.
The dot-short dashed line in each panel is the baryon 
accretion time for the result in ABN02.  The upper and lower diagonal
thin solid lines are the accretion times for constant accretion rates of $10^{-3}$ 
and $10^{-2}$~M$_\odot$/yr, respectively.
}
\label{fig.accrete.panel1}
\end{figure}

\begin{figure}
\begin{center}
\includegraphics[width=0.9\textwidth]{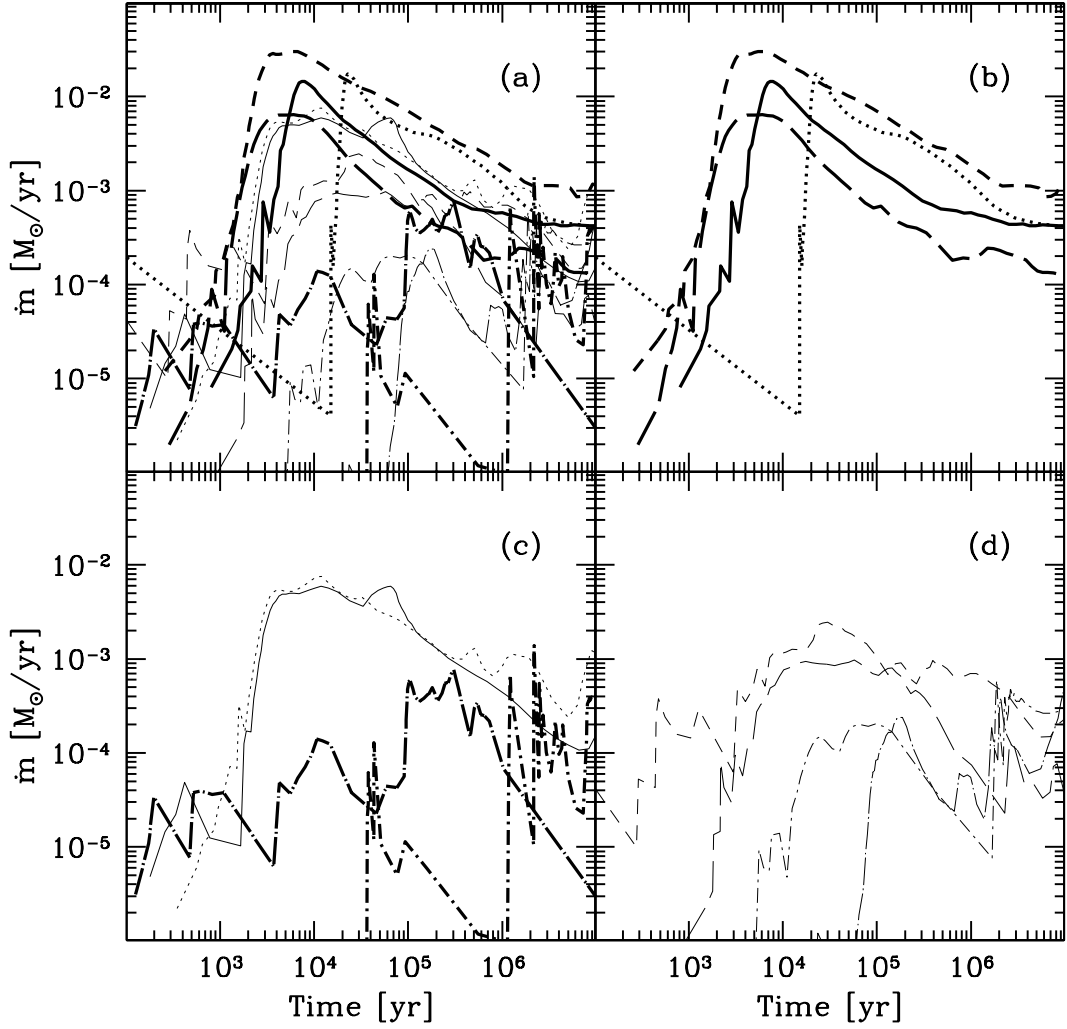}
\end{center}
\caption{
Mass-weighted, spherically averaged gas accretion rate as a function of 
time for all 12 cosmological random realizations.
Panel (a):  All simulations plotted
together.  Panel (b):  0.3 h$^{-1}$~Mpc box simulations.  Panel (c):
0.45 h$^{-1}$~Mpc box simulations.  Panel (d):  0.6 h$^{-1}$~Mpc box simulations. 
Line thicknesses and types are the same as in Figure~\ref{fig.comp.panel4}.
The accretion rate is defined as  $\dot{m} \equiv 4 \pi r^2 \rho(r) v(r)$, with $\rho(r)$ and 
$v(r)$ being the spherically-averaged baryon density and velocity as calculated on spherical shells,
and $v(r)$ defined as being positive towards the center of the halo.  The time is 
calculated by summing $\Delta M_b / \dot{m}$ in each shell starting at the
center of the halo and working outward, and plotting vs. $\dot{m}$.
}
\label{fig.accrete.panel1a}
\end{figure}

\begin{figure}
\begin{center}
\includegraphics[width=0.9\textwidth]{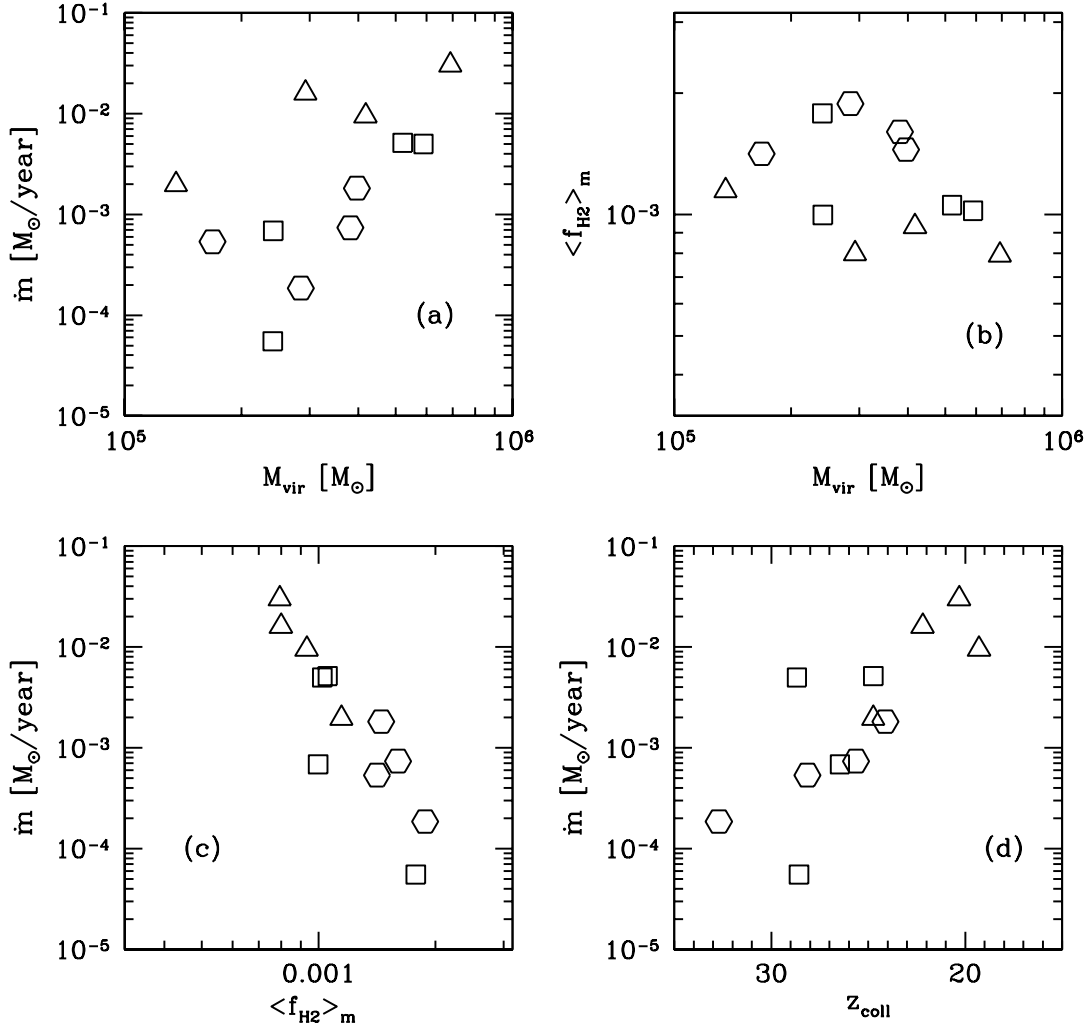}
\end{center}
\caption{
Panel (a):  Mass accretion rate ($\dot{m}$) vs. halo virial mass.  
Panel (b):  Molecular hydrogen fraction vs. halo virial mass.
Panel (c): Mass accretion rate vs. molecular hydrogen fraction.
Panel (d): Mass accretion rate ($\dot{m}$) vs. halo collapse redshift.  
The $H_2$ fraction is taken as an average over all gas within the virial radius.
All values of accretion rate are taken instantaneously at spherically-averaged radial
shells corresponding to an enclosed gas mass of 100 $M_\odot$.
In each plot, open triangles, squares and hexagons 
correspond to simulations with 0.3 h$^{-1}$~Mpc, 0.45 h$^{-1}$~Mpc and 0.6 h$^{-1}$~Mpc 
comoving box sizes, respectively.  
}
\label{fig.accrete.panel2}
\end{figure}

\begin{figure}
\begin{center}
\includegraphics[width=0.9\textwidth]{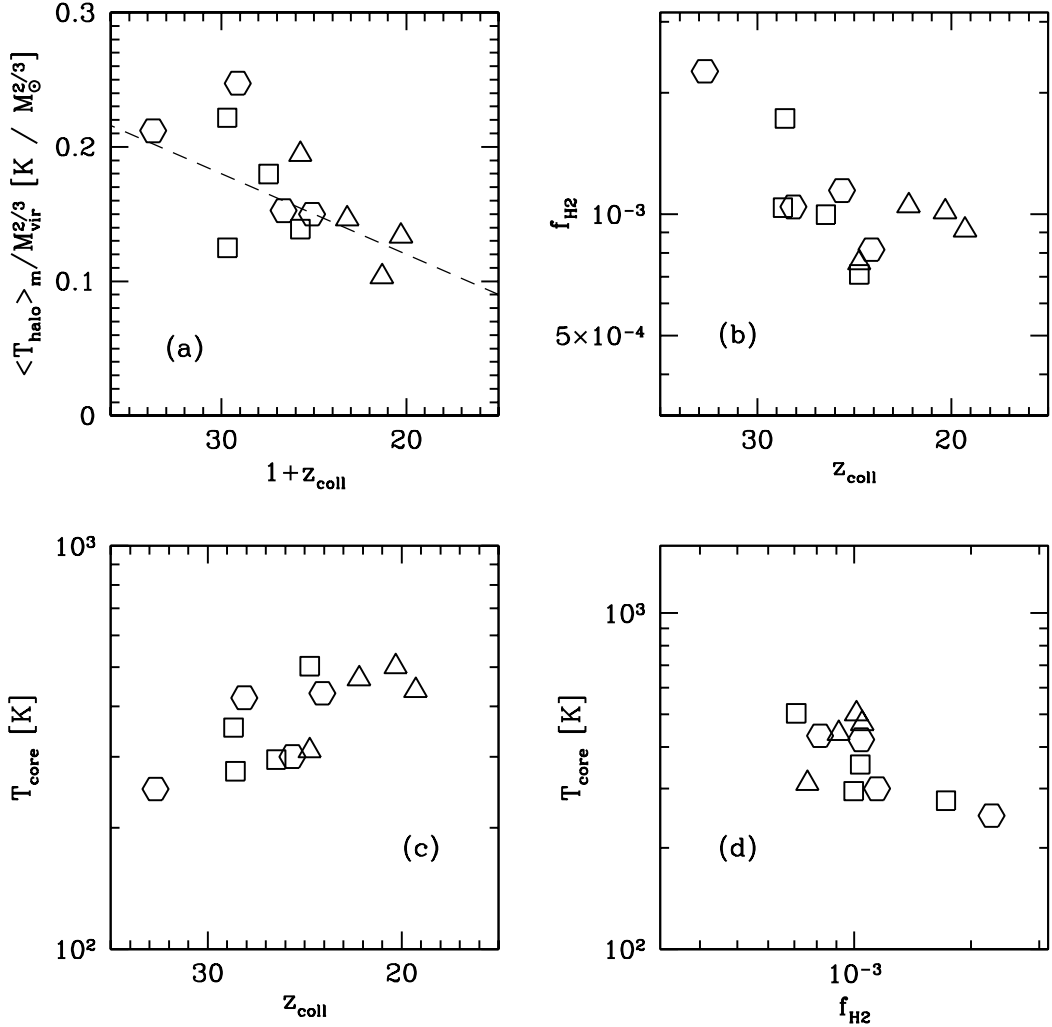}
\end{center}
\caption{
Panel (a):  $<T>_m/M_{vir}^{2/3}$ vs. $1+z_{coll}$ ($z_{coll}$ is the halo collapse redshift).
Panel (b):  Core $H_2$ fraction vs. collapse redshift.
Panel (c):  Core baryon temperature vs. collapse redshift.
Panel (d):  Core baryon temperature vs. $H_2$ fraction.
The dashed line in the top left panel presents the expected virial scaling of temperature with
mass and redshift, arbitrarily scaled.
Halo mean temperature is a mass-weighted average of all gas within the virial radius.  All ``core''
quantities are measured at the epoch of collapse at a proper
radius of 0.1 pc.
In each plot, open triangles, squares and hexagons 
correspond to simulations with 0.3 h$^{-1}$~Mpc, 0.45 h$^{-1}$~Mpc and 0.6 h$^{-1}$~Mpc 
comoving box sizes, respectively.  
}
\label{fig.accrete.panel3}
\end{figure}

\begin{figure}
\begin{center}
\includegraphics[width=0.9\textwidth]{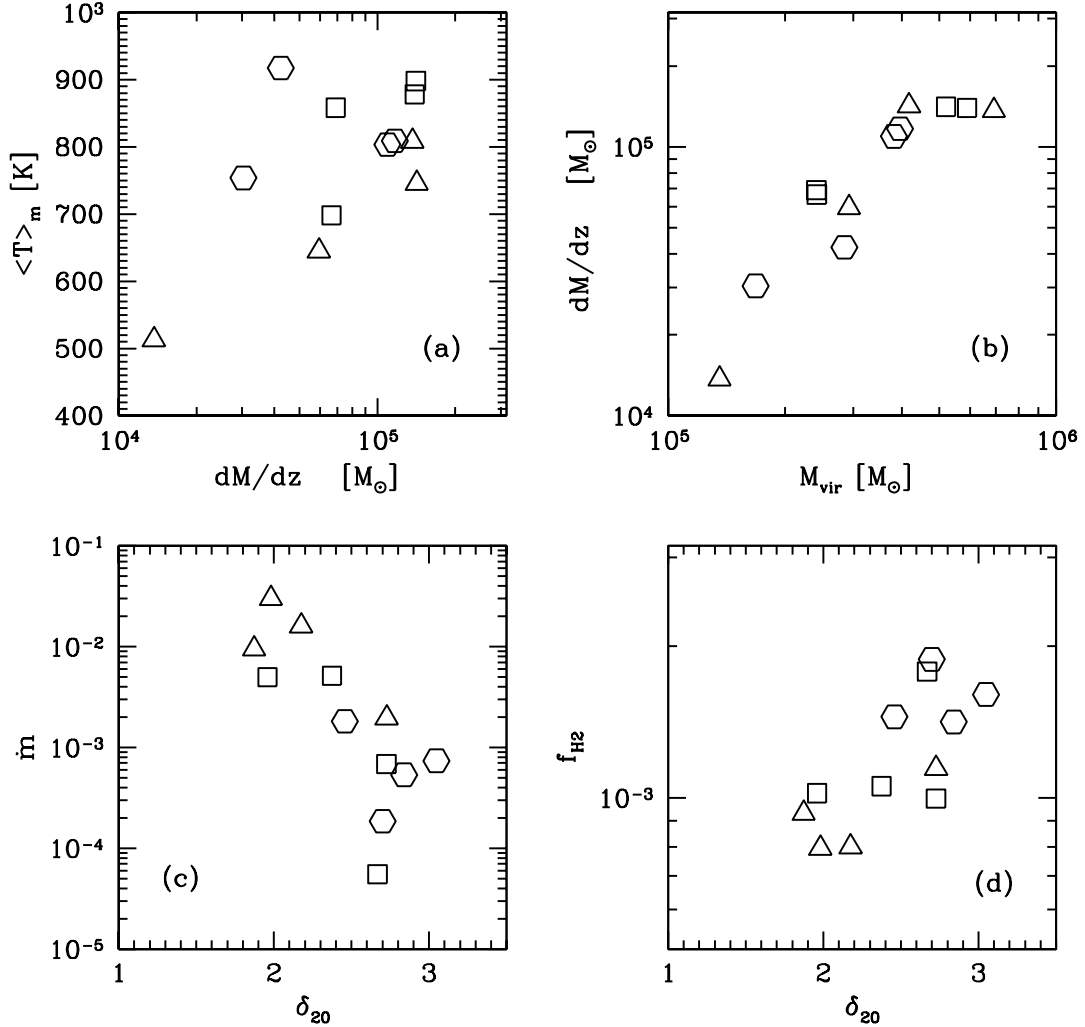}
\end{center}
\caption{
Panel (a):  Mean halo temperature vs. instantaneous halo mass accretion rate.
Panel (b):  Instantaneous halo mass accretion rate vs. halo virial mass.
Panel (c):  Core accretion rate vs. $\delta_{20}$.
Panel (d):  Halo mean molecular hydrogen fraction vs. $\delta_{20}$.
The ``core'' accretion rate in panel (c) is measured at the epoch of collapse at a proper
radius of 0.1 pc.
In panels (c) and (d), $\delta_{20}$ is defined as the mean density of the 
region within 20 virial radii of the halo (at the time of protostellar cloud collapse)
in units of $\Omega_m \rho_c$.  In each plot, open triangles, squares and hexagons 
correspond to simulations with 0.3 h$^{-1}$~Mpc, 0.45 h$^{-1}$~Mpc and 0.6 h$^{-1}$~Mpc 
comoving box sizes, respectively.  
}
\label{fig.accrete.panel4}
\end{figure}

Figures~\ref{fig.accrete.panel1} -- ~\ref{fig.accrete.panel4}
show baryon properties relating to the accretion rate onto the forming protostellar core.
Many of these properties are also detailed in Tables~\ref{table.meaninfo-1} 
and~\ref{table.meaninfo-2}.
Figure~\ref{fig.accrete.panel1} shows the mass-weighted, spherically averaged 
accretion time as a function of enclosed gas mass for 11 of the 12 cosmological
simulations.  There is a 2 order of magnitude scatter in accretion rates,
with a trend towards lower accretion rates (longer accretion times) in simulations
that have larger box sizes.  Figure~\ref{fig.accrete.panel1a} shows the mass-weighted,
spherically averaged accretion rate onto the protostellar core 
as a function of time for all twelve cosmological
simulations.  The time on the x-axis is calculated by summing up $\Delta M_b / \dot{m}$
on spherical shells, where $\Delta M_b$ is the amount of baryon gas and
$\dot{m}$ is the instantaneous gas accretion rate at that shell, 
starting at the center of the halo and assumes no feedback.
Although this is only a crude estimate of the true accretion rate, we see that all 
curves rise quickly to a maximum accretion rate at $t \sim 10^4$ year, and decline
thereafter, in agreement with the results of ~\markcite{2004NewA....9..353B}{Bromm} \& {Loeb} (2004)
and~\markcite{2006astro.ph..6106Y}{Yoshida} {et~al.} (2006).
\clearpage

\clearpage
Figure~\ref{fig.accrete.panel2} shows the mass accretion rate vs. halo virial mass,
molecular hydrogen fraction vs. halo virial mass, mass accretion rate vs.
molecular hydrogen fraction, and mass accretion rate vs. halo collapse redshift.
In all panels in this figure the molecular hydrogen fraction is taken as an average 
over all gas within the virial radius, and accretion rate values are taken
instantaneously at spherically-averaged radial shells corresponding to an
enclosed baryon mass of 100 M$_\odot$ (Note that the trends here do not change
significantly if this mass shell is varied by a factor of 10 in either direction,
so the results are quite robust).  This figure shows that there is a general 
trend of higher mass accretion rates with increasing halo mass.
though there is a great deal of scatter, and the variance is virial mass is not
that large.  There is no evidence for a comparable trend in the molecular hydrogen fraction 
with virial mass.  The bottom panels show that there is a strong relationship between 
the mass accretion rate and the mean molecular hydrogen fraction
of the halo and the redshift of collapse.
The accretion rate steadily decreases with increasing H$_2$ fraction, and increases
with decreasing redshift (increasing time).
\clearpage
\clearpage
The plots shown in Figure~\ref{fig.accrete.panel3} help to explain these trends.  This 
figure shows the mean halo temperature scaled by virial mass, 
core molecular hydrogen fraction, and core 
baryon temperature as a function of collapse redshift, and the core temperature as a
function of molecular hydrogen fraction.  All ``core'' quantities are measured at the 
epoch of collapse at a proper radius of 0.1 pc, though changing this value does not
change the results significantly.  There is a clear relationship between collapse
redshift and mean temperature, with temperature decreasing with decreasing 
collapse redshift (increasing time).  
This is explainable, as discussed previously, as a simple virial scaling: halo
masses at the time of protostellar core formation stay approximately constant
with redshift, and since the virial temperature scales as $T_{vir} \sim M_{vir}^{2/3} (1+z)$,
higher redshift halos have systematically higher temperatures.
The limiting reaction in the formation of molecular hydrogen at low ($\la 10^8$~cm$^{-3}$)
densities is the formation of H$^-$, which scales as $\sim T^{0.88}$.  This suggests
that halos whose gas is overall warmer (up to $\sim 1000$~K, where high temperatures
begin to collisionally dissociate H$_2$) more molecular hydrogen will be produced.  This is borne
out in the plot of core molecular hydrogen fraction vs. collapse redshift, where halos that form at
later times (and in overall cooler objects) have less molecular hydrogen.  The results of this are 
shown quite plainly in the plot of core temperature vs. collapse redshift, where objects that 
form later have higher core temperatures and hence higher accretion rates, since
$\dot{m} \sim T^{3/2}$.  This relationship is shown in a different way by the plot of core
temperature vs. core H$_2$ fraction in Figure~\ref{fig.accrete.panel3}, where there is a
clear correlation between decreasing temperature with increasing H$_2$ fraction.

\clearpage
\clearpage

Figure~\ref{fig.accrete.panel4} further explores the correlations shown in Figure~\ref{fig.accrete.panel3}.
Panel (a) of Figure~\ref{fig.accrete.panel4} plots the mean, mass-weighted halo temperature as a function
of instantaneous halo mass accretion rate.  The mass accretion rate is taken by calculating the difference
in virial mass between the halo at the time of protostellar core collapse and the virial mass of the most
massive progenitor at the last output time and dividing by the change in redshift.  There is a clear 
positive correlation between mean halo temperature and merger rate.  This is due to the hydrodynamic heating
of gas by mergers, as discussed by~\markcite{2003ApJ...592..645Y}{Yoshida} {et~al.} (2003) -- during halo mergers the gas is heated by 
shocks to approximately the virial temperature, and the gas in rapidly-growing halos does not have time to
cool to low temperatures before halo core collapse.  Panel (b) shows the halo mass accretion rate as a 
function of halo virial mass at the epoch of core collapse.  Halos which are undergoing rapid mergers 
show a strong tendency to be more massive, suggesting that a higher mass threshold must be reached before
the central density is high enough, and the $H_2$ formation and cooling rates are fast enough, for the core
to collapse.  We note that the apparent growth rate as a function of halo mass that we observe is higher
than that of~\markcite{2003ApJ...592..645Y}{Yoshida} {et~al.} (2003).  This may be due to our significantly higher mass and spatial 
resolution.

Panel (c) of Figure~\ref{fig.accrete.panel4} shows the instantaneous accretion rate onto the protostellar
core at the epoch of collapse as a function of $\delta_{20}$, which is defined as the mean overdensity of
the region within 20 virial radii of the collapsing halo.  $\delta_{20}$ is a proxy for local environment,
as used by~\markcite{2005MNRAS.363..379G}{Gao} {et~al.} (2005) -- regions with values of $\delta_{20} > 1$ are denser than an 
average region of the universe, with larger $\delta_{20}$ values corresponding to 
rarer regions.  There is a clear relationship
between protostellar core accretion rate and environment -- denser regions have lower accretion rates.
This is explained by panel (d), which plots the mean halo molecular hydrogen fraction as a function of
$\delta_{20}$.  Regions with higher values of $\delta_{20}$ tend to have higher molecular hydrogen fractions
and thus lower temperatures, which lead to lower accretion rates onto the protostellar core (as shown by 
Figure~\ref{fig.accrete.panel2} c).
\clearpage
\clearpage


\section{Discussion}\label{discuss}

In this paper we have explored several aspects of the formation of Population
III stars in a $\Lambda$CDM universe.  This section summarizes some of the processes
neglected in our calculations and also attempts to put some of the results in context.

The results presented in Section~\ref{compare-real} demonstrate that there is a 
great deal of scatter between the bulk halo properties such as overall virial mass,
collapse redshift, and mean halo temperature among the twelve simulations shown.
However, the final state of the density profile is extremely similar.  
This is due to the chemical and cooling properties
of the primordial gas.  The minimum temperature of the gas, which is determined by its chemical
composition, creates a density profile that goes roughly as $r^{-2}$ for any gas cloud which 
is only supported by thermal pressure.  This is for the gas contained in the 
halos out of which Population III stars form, so it is reasonable to expect consistent
density profiles on halo scales.

A detailed examination of the 
gas properties which may contribute significantly to the final Pop III star mass, such as
the core baryon temperature and accretion rate onto the forming primordial protostar, show
a significant amount of scatter.  This scatter can be attributed to variations in the molecular
hydrogen content of the halo on large scales brought on by differences in 
halo temperature as a result of varied merger rates between simulations, as suggested by 
Figures~\ref{fig.accrete.panel3} and~\ref{fig.accrete.panel4}.  Panel (a) of 
Figure~\ref{fig.accrete.panel3} shows a 
clear correlation between halo accretion rate and
temperature, suggesting that high halo merger rates cause an increase in halo temperature, 
which contributes to the scatter in mean temperatures for halos of a given mass.  There is
also a systematic relationship between halo collapse redshift and the mean temperature, with halos that
collapse at higher redshift having higher overall
halo temperatures and lower accretion rates onto the protostellar core.  
The higher temperatures due to redshift or merger 
rate result in somewhat larger molecular
hydrogen mass fractions, which cause the halo core to cool more rapidly during collapse.
Since the accretion onto the primordial protostar is primarily subsonic it is primarily controlled by
the local sound speed, with lower core temperatures directly resulting in lower accretion rates.  
We expect that as the gas collapses to higher densities ($n_H \gg 10^{12}~cm^{-3}$) the solutions will 
converge to the results found by~\markcite{omukai98}{Omukai} \& {Nishi} (1998), who find that
the late-time evolution of the gas can be approximated by a Larson-Penston similarity solution
(\markcite{larson69}{Larson} 1969;~\markcite{penston69}{Penston} 1969).  However, the 
lower-density cloud of gas around the protostar will not converge to a single solution, so the
star will still experience accretion with a range of histories and peak values.

One possible source of uncertainty in our understanding of the effect of redshift and environment
on the merger rates onto protostellar cores may come from the way in which halos are selected
for simulation in our calculations.  In each realization, we first run a 
dark matter-only calculation, and choose the most massive halo at $z=15$ for re-simulation with 
higher mass and spatial resolution and baryonic physics.  Selection of the most massive halo
in each calculation may result in a bias towards unusual halo formation environments and 
may skew our results.  Still, examination of Figure~\ref{fig.accrete.panel2} 
(d) shows significant scatter in accretion rates for halos which collapse at approximately the same
redshift, and suggests that both redshift and environment effects are important.  This issue
could possibly be clarified by performing an ensemble of 
simulations using halos chosen in a different manner.

After the onset of collapse, the evolution of the core of the halo, which has a mass of one thousand
solar masses, becomes effectively decoupled from the halo envelope since the time scales become
much shorter within the halo core.  This tells us that while the formation of the
initial primordial protostellar cloud is strongly coupled to the time scales associated with cosmological
structure formation, once the cloud has collapsed we can treat the core of the halo separately from the
rest of the calculation.  This decoupling will become highly useful when more detailed calculations of the
evolution of Population III protostars, including more complicated physics such as radiative transfer,
magnetohydrodynamics, and
protostellar accretion models, are performed, and will save us significant computational cost.

In Section~\ref{repstar}, we compare accretion times for a representative protostellar core
using accretion rates estimated with two different methods:  A simple calculation of 
instantaneous mass flux through shells using baryon density and radial velocity, and an 
estimate based solely on baryon temperature using the Shu isothermal collapse model.
These two values agree within a factor of two over a large range of mass scales
($10 \lesssim $~m$_{b, enc} \lesssim 10^3$~M$_\odot$).  The reason for this bears examination.
  As shown by \markcite{1977ApJ...214..488S}{Shu} (1977), as long as the
densities in a condensing molecular cloud core span several orders of magnitude before
a stage of dynamic instability is reached, the subsequent collapse properties
of the cloud should resemble those of an isothermal sphere.  The lack of characteristic
time and length scales results in a self-similar wave of infalling gas which propagates
outward at the speed of sound, resulting in an accretion rate that scales as
the cube of the sound speed.  This 
accretion rate can be derived in a more intuitive way by considering the properties 
of a cloud of gas with radius R and mass $M_{cl}$ which is marginally unstable.  The
accretion rate of this gas must be given (as an order of magnitude estimate) by
$\dot{m} \sim M_{cl}/t_{dyn}$, where $t_{dyn} = R/a$, where a is the characteristic
velocity associated with the cloud (the virial velocity).  If this cloud was originally
marginally supported against its self-gravity, then $a^2 \sim G M_{cl}/R$ (where G
is the gravitational constant), which can be
substituted into the expression for $\dot{m}$ to give $\dot{m} \sim a^3 / G$, 
independent of R.  In the case of this quasi-statically collapsing cloud, the virial
speed is comparable to the sound speed $c_s$, giving $\dot{m} \sim c^3/G$.  While the Shu
model assumes that the entire cloud is of a constant temperature, our calculations have a
varying temperature as a function of radius, and a radially-varying accretion rate based on
this temperature is an excellent fit.  This is reasonable because the isothermal collapse 
model assumes that the infall wave propagates at the local sound speed, assuming that the 
cloud is not supported by any other
means.  In this calculation we completely neglect the effects of magnetic fields,
and it can be seen from Figure~\ref{fig.rep.panel3} that rotation contributes an
insignificant amount of support to our representative protostellar cloud (as well as in 
the other 11 simulations discussed in this paper), resulting in gas pressure 
being the sole means of support
of the cloud.  One slight difference between the Shu model and our calculations is the 
presence of the dark matter halo, which results in a slightly steeper density profile 
($\rho \sim r^{-2.2}$), though the accretion rate result, being a function of sound speed,
is not changed significantly.

The observation that the rate of accretion onto the primordial protostar varies 
systematically as a function
of redshift, with halos that collapse at higher redshift having an overall 
lower accretion rate, has 
significant implications for
both reionization and metal enrichment of the early universe.  
There is no trend in the bottom right panel of Figure~\ref{fig.accrete.panel2} that
suggests an accretion rate ``floor'' exists at high redshifts.  However, one can 
make an estimate of the minimum possible accretion rate by observing that molecular hydrogen is only
effective at cooling the primordial gas down to approximately 200 Kelvin, which gives us an accretion
rate using the Shu model of $\dot{m}_{min} \simeq 8 \times 10^{-4}$~M$_\odot$/year, which is reasonably 
close to the accretion rates observed in the highest-redshift halos which have collapsed.
Though this implies convergence, it would be prudent to perform another suite of calculations 
at an even larger box size to be sure.  An absolute minimum accretion rate is set by the 
minimum temperature that primordial gas can attain,
which is effectively that of the cosmic microwave background, which scales as
T$_{cmb}(z) = 2.73 (1+z)~K$.  This gives a minimum accretion rate using the Shu model
of 

\begin{equation}
\dot{m}_{Shu,cmb} \simeq 1.2 \times 10^{-4} \left(\frac{1+z}{21}\right)^{3/2}~M_\odot/year
\end{equation}

\begin{figure}
\begin{center}
\includegraphics[width=0.8\textwidth]{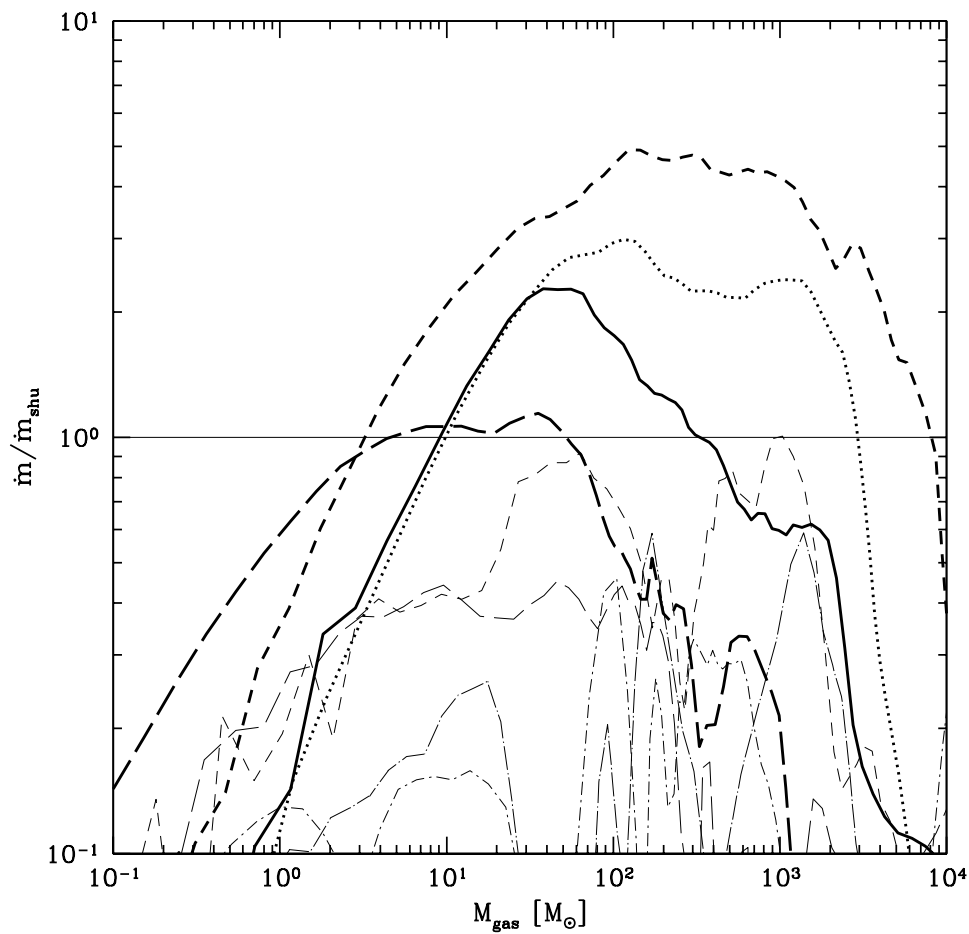}
\end{center}
\caption{
Ratio of accretion rate calculated using baryon density and velocity in mass-weighted, spherically
averaged shells vs. the accretion rate calculated using the Shu isothermal sphere model,  
plotted as a function of enclosed mass.  Lines weights and types correspond to those in 
Figure~\ref{fig.comp.panel4}, though for readability only simulations from the 0.3 and 0.6 $h^{-1}$ Mpc box
are plotted.
}
\label{fig.disc.mdotmshu}
\end{figure}

How reasonable is this estimate?  As shown in Figure~\ref{fig.rep.accrete}, the Shu
model does not agree perfectly with the accretion time results derived using baryon densities
and infall velocities, but does appear to provide a reasonable approximation for this one simulation.  
Examination
of our entire suite of simulations does not support this, however.
Figure~\ref{fig.disc.mdotmshu} plots the ratio of the accretion rate calculated from baryon density
and radial velocity with that estimated using the Shu model.  Lines are shown for only 8 of the
12 simulations for clarity.  This plot shows that simulated accretion rates depart significantly 
from the Shu model, which can be too high or low by a factor of several.  This 
suggests that the Shu model, while useful for a qualitative understanding of the physical system,
should not be considered to be quantitatively accurate for the purposes of estimating accretion rates.
\clearpage
\clearpage
The final masses of the stars in the calculations discussed in this work remains 
unclear.  These simulations lack the necessary radiative feedback processes which 
might halt accretion onto the protostar, making it impossible to accurately 
determine the main-sequence mass of the star.  However, rough bounds on the mass
range of these objects can be determined by examining Figure~\ref{fig.accrete.panel1}
and applying arguments similar to those used by ABN02.

A one solar mass object evolves far too slowly to halt accretion, particularly considering
the high rates at which mass is falling onto the star ($10^{-4}-10^{-2}$~M$_\odot$/year).  
This suggests that the star will have a mass which is
significantly higher than 1 M$_\odot$.  As discussed by~\markcite{2006astro.ph..2012B}{Beuther} {et~al.} (2006), low-mass 
protostars evolve on a timescale short compared to the Kelvin-Helmholtz timescale, whereas
the opposite is true for high-mass stars.  Thus, the Kelvin-Helmholtz timescale (plotted
in Figure~\ref{fig.accrete.panel1} using values of stellar radius and luminosity
from \markcite{2002A&A...382...28S}{Schaerer} (2002)) is a reasonable minimum mass scale for these objects, 
giving a lower bound to the stellar masses determined from these
calculations of $20-500$~M$_\odot$, depending on the simulation.

Massive stars can move onto the main sequence while still accreting rapidly.  However,
at high masses these objects have very short lifetimes ($\simeq 2 \times 10^6$ years for
a several hundred solar mass star) and this provides a reasonable upper limit to the 
mass of the star.  Comparing the stellar lifetime of these objects (as given
by~\markcite{2002A&A...382...28S}{Schaerer} (2002), we derive an upper mass limit of $\sim 40-2000$~M$_\odot$,
with individual stars having mass ranges that vary between $\sim 20-40$~M$_\odot$ and
$\sim 500-2000$~M$_\odot$, using this simple mass range estimate, with the
smaller mass objects forming systematically at higher redshifts.

This simple estimate fails to take into account essential physics.  As discussed by
~\markcite{2003ApJ...589..677O}{Omukai} \& {Palla} (2003) and \markcite{TanMcKee2004}{Tan} \& {McKee} (2004), radiative feedback from
the evolving protostar will strongly affect the dynamics of the accreting gas.
\markcite{TanMcKee2004}{Tan} \& {McKee} (2004) and McKee \& Tan (2006, in prep.) use a combination of analytic
and semi-analytic models of the evolving system, and show that feedback does not limit
accretion until the star has reached $\simeq 30$~M$_\odot$, and suggest that ionization
creates an HII region that limits growth of stars to masses less than $\sim 100$~M$_\odot$.
They assume a very high (and time varying) accretion rate compared to the mean values
observed in our suite of simulations, and it is unclear how varying this would affect
the final mass of the object.

\markcite{2003ApJ...589..677O}{Omukai} \& {Palla} (2003) perform 1D, spherically-symmetric calculations of the 
evolution of a Population III protostar undergoing rapid accretion.  They include
radiative feedback effects, and find that there is a critical mass accretion rate,
$\dot{m}_{crit} \simeq 4 \times 10^{-3}$~M$_\odot$/year, below which objects can accrete to
masses much greater than 100~M$_\odot$ (essentially up to the limit of gas available 
in the halo core and the main sequence lifetime of the forming object), and below which
the maximum mass decreases as the accretion rate increases.  They find that if they assume
 a time-varying accretion rate that stars at high values and decreases rapidly with time, 
the evolution of the star follows the $\dot{m} \leq \dot{m}_{crit}$ path, and accretion can
 continue until the end of the star's main sequence lifetime.

As can be seen from this discussion, it seems that the arguments for mass range given
by ABN02 are overly simplistic.  However, the models discussed above do not completely
resolve the issue of Population III IMF -- Tan \& McKee use a single accretion rate
for their work (which includes multi-dimensional effects), and Omukai \& Palla
perform spherically-symmetric calculations which may miss essential accretion shutoff 
physics.  Clearly, the relationship between accretion rates and final stellar mass
cannot be stated with any certainty, and will have to be addressed using multi-dimensional
numerical simulations which include protostellar evolution and radiative feedback.
However, if one assumes that the main-sequence lifetime of an accreting object is
a reasonable proxy for the maximum stellar mass, it seems that Population III stars
that form at higher redshifts should generally have smaller main-sequence masses.

If in fact a lower overall accretion rate results in a less massive population of stars, 
these objects will be much less effective at ionizing 
the intergalactic medium (since they produce overall fewer UV photons per baryon) and will produce a 
completely different nucleosynthetic signature.  This may be in accord with the
WMAP Year III measurements of the polarization of the cosmic microwave background, which
suggests that the universe was significantly reionized by $z \simeq 10$~\markcite{2006astro.ph..3450P}({Page} {et~al.} 2006).
Models show that the observed level of polarization does not require a large contribution
to reionization from Population III stars.  Additionally, \markcite{2006ApJ...641....1T}{Tumlinson} (2006) uses a
semianalytical model constrained by the metallicity distributions of Population II stars 
in the galactic halo, relative abundances of ultra metal poor halo stars, and estimates 
of the ionizing photon budget required to reionize the IGM, and suggests that Population
III stars have an IMF which has a mean mass of $8-42$~M$_\odot$. 

One factor that has been completely ignored in this work is the effect of a soft UV
background.  Massive primordial stars are copious emitters of light in the Lyman-Werner
band ($11.18-13.6$ eV).  Atomic hydrogen is transparent to these photons, which 
photodissociate molecular hydrogen.  This effect, explored by \markcite{2001ApJ...548..509M}{Machacek} {et~al.} (2001)
and others, increases the threshold mass of halos which form Pop III stars and may also
result in an increase in temperatures in the halo cores, causing an increase in the
overall accretion rate onto the forming protostars.  We explore this issue in detail
in a forthcoming paper (O'Shea \& Norman 2006, in preparation).  
A semianalytical examination of this effect was made by 
\markcite{2005ApJ...629..615W}{Wise} \& {Abel} (2005), who use Press-Schechter 
models of Population III star formation to predict a slowly rising Lyman-Werner background which provides
some support to this idea.  This suggests that further calculations including larger simulation volumes as
well as a soft UV background will be necessary to make a definitive statement about the most common accretion
rates.  Additionally, these calculations completely neglect the mode of primordial star formation that takes 
place in halos whose virial temperatures are above $10^4$ K.  Cooling in these systems is dominated
by atomic hydrogen line emission and, particularly in the presence of a strong soft UV background, may result
in a much larger amount of cold gas distributed in a different manner than in the systems simulated in this
work, which have mean virial masses of a few times $10^5$~M$_\odot$ and virial temperatures of around 1000 K.
These systems have been examined numerically by \markcite{2003ApJ...596...34B}{Bromm} \& {Loeb} (2003), but further
work is necessary to explore the properties of these objects.

Figure~\ref{fig.rep.panel3} in Section~\ref{repstar} shows clear signs of angular momentum evolution in the collapsing
protostellar core.  Angular momentum is less of an issue in Population III star formation than in galactic
star formation - the collapsing cosmological halo out of which the protostellar core forms has very little angular 
momentum from the outset (with a spin parameter of $\lambda \simeq 0.05$), and violent relaxation during virialization
results in an angular momentum distribution $l \sim r$, so circular velocities are fairly constant throughout the halo.
The $\sim 1000$~M$_\odot$ halo core out of which the protostar forms is not rotationally supported -- the gas is
essentially completely held up by thermal pressure, and the small amount of angular momentum that is actually transported
is not a critical factor in the cloud core's collapse.  This angular momentum redistribution, as discussed in 
Section~\ref{repstar}, appears to be due to a combination of angular momentum segregation and turbulent transport.
 Similar results were reported by~\markcite{2006astro.ph..6106Y}{Yoshida} {et~al.} (2006). 
A detailed analysis of angular momentum transport issues will be discussed in the second paper in this series
(O'Shea, Norman \& Li 2006, in preparation).

Though the collapsing protostellar cloud is not rotationally supported by the time the central density
reaches $n_H \simeq 10^{15}$~cm$^{-3}$, it is inevitable that at some point a protostar with a 
rotationally-supported accretion disk will form, as discussed in~\markcite{TanMcKee2004}{Tan} \& {McKee} (2004).  Unfortunately,
this takes place at high densities, where the optically-thin assumption used in the Enzo cooling algorithms
breaks down, which necessitates either an extension of the cooling model~\markcite{ripamonti04}({Ripamonti} \& {Abel} 2004) or the 
use of a radiation hydrodynamics code such as ZEUS-MP~\markcite{2005astro.ph.11545H}({Hayes} {et~al.} 2005).  This 
issue will be approached in later work.

The primordial chemistry model used in these calculations ignores the effects of deuterium, lithium, and the
various molecules they form between these elements and ordinary hydrogen.  Deuterium and lithium 
have been shown to be unimportant in the temperature and density regimes that we have examined in this
work~\markcite{1998A&A...335..403G}({Galli} \& {Palla} 1998).  However, it is possible that they may be relevant in other situations of importance to 
Population III star formation -- in particular, regions which have been ionized to very high electron
fractions may experience significant cooling from the HD molecule, which has a low excitation energy and 
forms preferentially compared to $H_2$ due to chemical fractionation.  This is particular true
at low temperatures ($T < 150$~K) where essentially all deuterium is converted to HD.  Cooling due to
the HD molecule 
has the potential to cool gas down to approximately the temperature of the cosmic microwave background, which 
scales with redshift as $T_{cmb}(z) = 2.73 (1+z)$~K~\markcite{2000MNRAS.314..753F,
2002P&SS...50.1197G,2005MNRAS.361..850L}({Flower} {et~al.} 2000; {Galli} \& {Palla} 2002; {Lipovka}, {N{\'u}{\~n}ez-L{\'o}pez}, \&  {Avila-Reese} 2005).  This gives a minimum baryon temperature of 
approximately $55$ Kelvin
at $z=20$ and could further reduce the minimum accretion rate onto a primordial protostar.
Lithium, while in principle an effective coolant when in LiH, is safely ignored, since only a tiny fraction of 
lithium is converted into molecular form~\markcite{mizusawa2005}({Mizusawa} {et~al.} 2005).
A final chemical process which is omitted from our calculation is heating caused by
molecular hydrogen formation at high densities ($n_H > 10^8$~cm$^{-3}$), the inclusion of
which which may result in
differences in the temperature evolution of the gas in the highest density regimes
considered here.  This does not affect the main conclusions of our paper, as significant
differences between halos can be observed at densities much lower than $10^8 cm^{-3}$.

The effects of magnetic fields are completely ignored in the simulations discussed in this work.  We can justify 
this by examining the magnetic field necessary to delay the collapse of the halo core.  If one assumes that
the halo core can be represented reasonably well by an isothermal sphere of constant density (which is reasonable at 
the onset of halo collapse), we can use the virial theorem to estimate the strength of the magnetic field which is
necessary to support the collapse of the halo against gravity.  Assuming flux freezing and a uniform magnetic field,
a magnetically critical isothermal sphere has a mass-to-flux ratio of

\begin{equation}
\frac{M_{cl}}{\Phi_B} = \frac{1}{\sqrt{31 G}}
\label{eqn-isosphere}
\end{equation}

Where M$_{cl}$ is the mass of the halo, $\Phi_b = \pi R_{cl}^2 B_{cl}$ is the magnetic flux in the cloud 
(with R$_{cl}$ and B$_{cl}$ being the cloud radius and magnetic field strength, respectively), and G
is the gravitational constant.  Reasonable values for M$_{cl}$ and R$_{cl}$ are $\simeq 2 \times 10^3$~M$_\odot$ 
and 4 parsecs, respectively, which gives us a value of the magnetic field of B$_{cl} = 1.21 \times 10^{-5}$ G.  
The mean density of the cloud is n$_{cl} \simeq 300$~cm$^{-3}$ and the mean density of the universe at $z=18$ 
(the redshift that our cloud collapses) is $\simeq 0.003$~cm$^{-3}$, so if we assume a spherical collapse from
the mean cosmic density assuming flux freezing, we see that
the ratio of the magnetic field in the cloud to the mean universal magnetic field is

\begin{equation}
\frac{B_{cl}}{B_{igm}} = \left( \frac{n_{cl}}{n_{igm}} \right)^{2/3}
\label{eqn-fluxfreeze}
\end{equation}

This gives us a mean magnetic field of $B_{IGM} \simeq 3.5 \times 10^{-9}$~G at $z \simeq 18$.  Since there
are no known objects that may produce magnetic fields between recombination ($z \sim 1100$) and the
epoch of Pop III star formation, and the strength of the magnetic field scales with the expansion of the universe as 
$(1+z)^2$, we estimate that in order for magnetic fields to be dynamically important in the formation of
Population III stars the magnetic field strength at recombination must be $B_{rec} \sim 10^{-5}$ G.
The current observational upper limit to magnetic field strength at recombination (albeit at large scales) is 
B $\leq$ 3 x 10$^{-8}$ G as measured at the present epoch~\markcite{2000PhRvL..85..700J}({Jedamzik}, {Katalini{\'c}}, \&  {Olinto} 2000), which 
corresponds to a magnetic field at 
recombination of approximately $4 \times 10^{-2}$ G.  This is three orders of magnitude higher than
needed to be dynamically relevant for Population III star formation!  
The only known mechanism that could generate coherent magnetic fields with strengths of this scale is
a discontinuous QCD or electroweak phase transition.  However, the mechanisms involved are highly 
speculative and predictions of the possible magnetic field strengths are unreliable~\markcite{1997PhRvD..55.4582S}({Sigl}, {Olinto}, \&  {Jedamzik} 1997).
Currently, the most
plausible mechanisms for creating magnetic fields at recombination suggest that field strengths are on
the order of $10^{-23}$ G at
recombination~\markcite{2005PhRvD..71d3502M}({Matarrese} {et~al.} 2005).  Given the observational uncertainty, it seems reasonable to ignore this
effect, though future simulations will certainly include magnetic fields with a variety of field strengths
and physical scales..

Assuming that the magnetic field at that epoch was strong enough to be dynamically important, we can calculate the effect
that this has on the collapse of the star.  Due to the relatively high electron fraction in the halo core ($f_e \sim 10^{-6}$
at $r \simeq 1$ pc; see Figure~\ref{fig.rep.panel2} (b)) the assumption of flux freezing in the magnetic field at 
scales of $\sim 1$ pc is valid~\markcite{makisusa04}({Maki} \& {Susa} 2004).  However, the
electron fraction rapidly declines at smaller radii, reaching $f_e \sim 10^{-10}$ at $\sim 200$ AU, and this
assumption may no longer be relevant.
 Magnetic fields couple to charged particles (electrons and ions),
and these charged particles interact with the neutral medium.  At high levels of 
ionization, collisions between charged and
 neutral particles are frequent, implying that the magnetic field is strongly coupled to the gas.  However, at low
levels of ionization there are few charged particles, and the coupling with the neutral gas is weak.  In an object
that is subject to a gravitational acceleration this produces a relative drift of charged and neutral particles 
which allows the neutral gas to decouple from the magnetic field.  This effect is known as ``ambipolar diffusion,''
and is believed to be an extremely important process in galactic star 
formation~\markcite{1991ApJ...371..296M}({Mouschovias} \& {Morton} 1991).  
The retardation effect that ambipolar
diffusion may have on the collapse of the halo core can be estimated by examining the relative timescales of ambipolar 
diffusion and halo collapse.  The ambipolar diffusion timescale can be estimated as

\begin{equation}
\tau_{AD} = \frac{L}{v_D} \simeq 2 \times 10^6~\frac{x_i}{10^{-7}}~years
\label{eqn-ambipolar}
\end{equation}

Where L and $v_D$ are a characteristic length scale and the neutral-ion relative drift velocity, respectively, and $x_i$ is
the overall ionization fraction.  A proxy
for the halo collapse time scale is the free fall time, which for a spherical system is

\begin{equation}
\tau_{ff} = \left( \frac{3 \pi}{32 G \rho} \right)^{1/2} \simeq \frac{5 \times 10^7}{n^{1/2}} years
\label{eqn-dyntime}
\end{equation}

where n is the particle number density in particles per cubic centimeter and G is the 
gravitational constant.  The relevance of ambipolar diffusion
can be estimated by taking the ratio of these two quantities, which is known as the ``collapse retardation time,''
$\nu_{ff}$.  Substituting in equations~\ref{eqn-ambipolar} and~\ref{eqn-dyntime}, we see that

\begin{equation}
\nu_{ff} \equiv \frac{\tau_{AD}}{\tau_{ff} } \simeq 4 \times 10^{5} x_i n^{1/2}
\label{eqn-collretard}
\end{equation}

Examination of figures~\ref{fig.rep.panel1} and~\ref{fig.rep.panel2} show that at the final timestep
in the calculation, the number density can be fitted by a power law and is roughly 
$n(r) \simeq 10^3~(r/pc)^{-2}$~cm$^{-3}$ while the ionization fraction scales roughly as 
$x_i(r) \simeq 10^{-6}~(r/pc)$.  Plugging these into equation~\ref{eqn-collretard} shows that 
$\nu_{ff} \simeq 13$ is constant with radius.  This is only a crude approximation, since the free fall time
really should depend on the mean number density instead of the number density at a given radius.  However, 
considering the rapid falloff of density, n(r) is a reasonable approximation of n -- strictly speaking, for a
cloud with a density profile that scales as $r^{-2}$ over many orders of magnitude in radius, 
the mean density is equal to $3~n(r)$, so our estimate
of the free fall time is too high by a factor of $\sqrt{3}$.  Plugging this in to the equation, we get that
$\nu_{ff} \simeq 23$ everywhere, which indicates significant delay in collapse with respect to the free fall time.
However, the relevant time scale in this case is more appropriately the quasistatic collapse time, which is 
approximated as $\tau_{qs} \simeq L/v_{r}$.  Figure~\ref{fig.rep.panel3} shows that the mean radial velocity
at the scales of interest ($\sim 2-3$ parsecs) is roughly 0.5 km/s.  This corresponds to $\tau_{qs} \sim 4 \times 10^{6}$
and scales linearly with the radius.  Comparison with the ambipolar diffusion time scale shows that $\tau_{AD}$ and
$\tau_{qs}$ are within a factor of two of each other, which suggests that the presence of a magnetic field would not
significantly impede the collapse of the halo core for the quasistatic collapse case.  It is interesting to note that
~\markcite{2003MNRAS.341.1272F}{Flower} \& {Pineau des For{\^e}ts} (2003) reach a similar conclusion using a significantly different method.

There are several effects which we have neglected in these calculations which 
may affect our results.  The box sizes of the calculations presented in this 
paper are relatively small (300 -- 600 kpc/h comoving).  \markcite{2004ApJ...609..474B}{Barkana} \& {Loeb} (2004)
show that finite box sizes can have significant effects on statistical properties 
of halo formation, resulting in an overall bias towards under-sampling of the mass
function and late halo formation times.  However, our calculations are not intended
to provide a statistically accurate sampling of the mass function of dark matter
halos.  Rather, the intent is to gain an understanding of the first halo to undergo
baryonic core collapse in a simulation volume.  An obvious extension of this work
would be to simulate a much larger simulation volume so that cosmic variance and
possible environmental effects are considered (as discussed by Barkana \& Loeb), and then
simulate multiple halos within this volume.  An example of this technique
is presented by~\markcite{2005MNRAS.363..393R}{Reed} {et~al.} (2005),
who simulate the formation of the first star-forming halo in a simulation volume which is
$497$~h$^{-1}$ Mpc on a side.  They find that the first halo of mass $1.2 \times 10^5$~h$^{-1}$~M$_\odot$
forms by $z=49$, and would be able to collapse by $z \simeq 47$, when the halo has reached
a mass of $\simeq 2.4  \times 10^5$~h$^{-1}$~M$_\odot$, which agrees reasonably well with
estimates for a minimum halo mass to collapse and form a Population III star, as determined
in this work.  The virial temperature of this halo would be a factor of 2.4 higher than that of
a halo which virializes at $z=20$, and the $H_2$ production rate would be approximately
2.2 times higher.  Compared to a halo that forms at $z=32$ -- the highest redshift halo we 
examine -- the differences in virial temperature and $H_2$ formation rate are 1.45 
and 1.39, respectively.  Though somewhat simplistic, these arguments suggest that the range
of halo properties explored in our set of simulations may encompass most of
the expected range of behavior.  Though the Reed et al. simulations do not include
baryonic physics, such a calculation is technically possible using the Enzo code, and is a 
logical extension of our work.

Another important consideration is the generation of cosmological initial conditions.
\markcite{2006astro.ph..1233H}{Heitmann} {et~al.} (2006) explore
the evolution of the halo mass function at high redshifts, and conclude that
the overall mass function is suppressed in simulations whose initial conditions
are generated at too late of a redshift.  They suggest that a starting
simulation redshift of $z \simeq 500$ would be appropriate for the small boxes used in 
the calculations discussed in this paper. The magnitude of error involved in simulations s
tarting later is
not quantified.  Since our focus is only on the most massive halo to form in the
simulation volume, the magnitude of the bias in our results by starting at $z=100$ 
is unclear.

A fundamental assumption that we make concerns the validity of the cold dark matter
model at mass and spatial scales comparable to, or greater than, the masses of
the halos which form Population III stars.  Reliable measurements of the cosmological 
power
spectrum only exist down to the length scales corresponding to the Lyman-$\alpha$ 
forests, which is composed of structures that are much more massive than 
$\sim 10^6$~M$_\odot$ \markcite{2005ApJ...635..761M,2005MNRAS.361...70J,2006astro.ph..3449S}({McDonald} {et~al.} 2005; {Jena} {et~al.} 2005; {Spergel} {et~al.} 2006).
If the power spectrum is suppressed at small scales, this could have significant implications
for the formation epoch and properties of Population III stars~\markcite{2006oshea_wdm}({O'Shea} \& {Norman} 2006).
Additionally, a running power spectrum or significant changes in $\sigma_8$ or the
primordial power spectrum index (as suggested by the WMAP Year III results)
may have significant effects, which should be explored.

\section{Summary}\label{summary}

In this paper we have performed a suite of twelve high dynamical range simulations
of the formation of Population III stars in a $\Lambda$CDM universe, with the intent to
further constrain the variation of accretion rates onto Population III protostellar 
cores, a parameter that is of crucial importance to the mass scale of Population 
III stars.  Our simulations span a range of box sizes and are each
an independent cosmological realization.  Our principal results are as follows:

1.  Our results are quantitatively and qualitatively similar to those of 
ABN02, which was a single realization performed in an $\Omega_m = 1$
SCDM universe.  This suggests that the scenario presented in that work is 
generally correct.  We note that~\markcite{2006astro.ph..6106Y}{Yoshida} {et~al.} (2006) have come to
the same conclusion using different numerical techniques.

2.  We observe no signs of fragmentation in the collapsing protostellar core
in any of our 12 simulations, suggesting that the paradigm of a massive
Population III star forming in isolation in a halo of mass $\sim 10^5 - 10^6$~M$_\odot$
is robust.

3.  The virial masses of halos which form Population III protostars do not evolve
significantly with redshift in calculations without a soft UV background, and a rough
threshold mass of $\simeq 1.5 \times 10^5$~M$_\odot$ below which protostellar cores do
not form is observed.

4.  Other bulk halo properties, such as spin parameter and angle of separation between
gas and dark matter angular momentum vectors, are consistent with previous numerical 
simulations of much larger cosmological halos, and do not appear to evolve with redshift.

5.  Our suite of calculations suggest a clear relationship between halo collapse redshift 
and environment and the observed accretion rate onto the forming protostellar core, 
with lower accretion rates seen at 
higher redshift and in halos which form in significantly overdense regions of the universe. 
This is explained by cooler gas in the cores of these halos, and
is the result of higher molecular hydrogen fractions from warmer overall gas due to the virial scaling of
halo temperature with redshift or due to halo's merger history.

\acknowledgments{
BWO would like to thank Tom Abel, Greg Bryan, Simon Glover, 
Alex Heger, Savvas Koushiappas, Jonathan Tan and Matthew Turk
for useful discussion, and to the Center for Astrophysics and 
Space Sciences at the University of California in San Diego for
hosting him during the completion of a portion of this project.  
MLN would like to thank the Institute for Advanced Study at 
Princeton for their support and hospitality during the period 
of March--June 2006, during which this manuscript was finalized.
BWO and MLN would like to acknowledge the Institute for Nuclear 
Theory at the University of Washington for their support and 
hospitality.  Special thanks to Jonathan Tan for providing BWO 
with a copy of an as-yet-unpublished manuscript. This work has 
been supported in part by NASA grant NAG5-12140 and NSF grant 
AST-0307690.  This work was carried out in part under the 
auspices of the National Nuclear Security Administration of the
U.S. Department of Energy at Los Alamos National Laboratory under 
Contract No. DE-AC52-06NA25396.  The simulations were performed 
at SDSC and NCSA with computing time provided by NRAC allocation 
MCA98N020.  
}

\bibliography{}  

\end{document}